\documentstyle[preprint,aps,fleqn,amssymb]{revtex} 

\newfont{\bi}{cmbxti10 scaled\magstep2}
\begin{document}
\draft

\title{A scaling hypothesis for corrections to total energy and stress
in plane-wave based {\bi   ab initio} calculations}
\author{G.-M. Rignanese, Ph. Ghosez, J.-C. Charlier, J.-P. Michenaud,
and X. Gonze.}
\address{Unit\'e de Physico-Chimie et de Physique des
Mat\'eriaux,
Universit\'e Catholique de Louvain,\\
1 Place Croix du Sud, B-1348 Louvain-la-Neuve, Belgium}
\date{\today}
\maketitle

\begin{abstract}
We present a new technique aimed at preventing plane-wave based
total energy and stress calculations
from the effect of abrupt changes in basis set size.
This scheme relies on the interpolation of energy
as a function of the number of plane waves,
and on a scaling hypothesis that allows to perform the interpolation
for a unique reference volume.
{}From a theoretical point of view, the new method is compared to those
already proposed in the literature, and its more rigorous derivation
is emphasized.
{}From a practical point of view, we illustrate the importance of the
correction
on different materials (Si, BaTiO$_3$, and He) corresponding to different
types of bonding, and to different $k$-point samplings and cut-off energies.
Then, we compare the different approaches for the calculation of
$a_0$, $B_0$, and $B_{0}^{'}$ in bulk silicon.
\end{abstract}



\section*{Introduction}

Since the early eighties, the Local Density Appoximation (LDA)
\cite{Kohn-Sham65}
to Density Functional Theory (DFT)
\cite{Hohenberg-Kohn64}
has proven to be a choice tool to obtain reliable
total energies of solids.
Application of the total energy method to a solid of
given (or assumed) crystal structure usually begins by
the determination of the static equilibrium properties,
like the lattice constant $a_0$, the bulk modulus $B_0$, and possibly
its pressure derivative $B_{0}^{'}$.
This determination is carried out before evaluation of other
properties such as phonon spectra.
Considerable help in this task arises from the stress theorem, derived by
Nielsen and Martin
\cite{Nielsen-Martin83,Nielsen-Martin85a,Nielsen-Martin85b}
that complements a calculation of the total energy, by giving the six
components
of the macroscopic stress tensor $\sigma_{\alpha \beta}$,
at practically no cost.

For simplicity, we first consider a cubic structure
(for which there is only one lattice constant).
The hydrostatic pressure is related to the total energy and the volume $V$ by~:
\begin{equation}
P=P(V)=-\frac {\sigma_{11}+\sigma_{22}+\sigma_{33}} {3}
=-\frac {\partial E_{tot}} {\partial V}
\,.
\label{eq:pressdef}
\end{equation}

In this case, the static equilibrium properties can easily be determined
either from the curve of energy $E_{tot}$ versus lattice constant $a$
(or, in an equivalent way, versus volume $V$)
or from that of pressure $P$ versus lattice constant $a$.
These curves can be obtained by fitting a polynomial to the values of
$E_{tot}$ and $P$ calculated for several values of the lattice constant.

We now describe the problems that arise when a plane-wave based method
is used for total energy and pressure calculation.

Considering a periodic cell, the electronic wave functions should be
expanded in terms of an {\it infinite} set of plane waves
(Fourier series) at each of an {\it infinite} set of $k$-points
in the Brillouin zone.
The following two approximations are introduced.
Firstly, a small number of carefully chosen $k$-points (special points)
are used to sample the Brillouin zone
\cite{Monkhorst-Pack76}.
Secondly, the wavefunctions at each $k$-point are expanded in terms of
a {\it finite} basis set of plane waves $e^{i(k+G)r}$ such that their kinetic
energy is smaller than a fixed cut-off energy $E_{cut}$
(Hartree atomic units are used throughout the paper)~:
\begin{equation}
\frac {1} {2} \left| k+G \right|^2 < E_{cut}
\,.
\label{eq:defpwset}
\end{equation}

In principle, it is possible to approach energy convergence by
augmenting the number of special $k$-points and the cut-off energy (which
corresponds to increasing the size of the basis set).
However, working with large number of special points and plane waves
requires a huge amount CPU time.
In order to perform calculations on larger and more complex systems,
one aims at using smaller plane-wave basis sets at each
$k$-point without reducing the accuracy of the calculation.

{\it Differences} in the total energy of systems with the same
unit cell are known to be accurately calculated for a number of plane waves
and special $k$-points smaller than those required to ensure
convergence of the energy, provided that {\it identical} plane-wave
basis sets are used for each calculation \cite{Cheng88}.
The error due to the truncation of the Fourier series is systematic
and cancels out.
One could expect $a_0$, $B_0$, and $B_{0}^{'}$ to have similar properties~:
part of the systematic error in $E_{tot}$ should cancel.
However, when computing energy differences between systems of varying
size, it is impossible to use {\it identical} plane-wave basis sets due
to the periodic boundary conditions imposed on the edge of the cell
whose dimensions change with the volume.
Instead we must choose either to use a constant number of plane waves
$\overline{N}_{PW}$ \cite{Note_kpoints} in the basis set with $E_{cut}$
depending upon the volume,
or a constant cut-off energy for determining the plane wave basis set.

The $E_{tot}$ curve for constant $\overline{N}_{PW}$ is always very smooth
since the set of plane waves scales smoothly with the lattice constant.
Unfortunately, relying on this curve leads to a systematic underestimation
of the lattice constant, as discussed in previous papers \cite{Dacosta86}.
On the other hand, constant $E_{cut}$ corresponds
to a constant resolution in real space,
which gives less biased results \cite{Dacosta86};
but, the curve for constant $E_{cut}$ is ragged.
One gets a set of disconnected micro-curves
(a micro-curve being a continuous segment of $E_{tot}$ or $P$)
as shown in Fig.~\ref{fig:segments},
whereas the experimental curves are perfectly continuous.
This stair-like variation is due to the discontinuous increase of the
number of plane waves accompanying the unit cell volume changes.
Every time new $(k+G)$ vectors are added to the basis set, $E_{tot}$
discontinuously decreases due to the added variational freedom
in the wave function.
The $P$ curve also presents discontinuities for the same reasons.
Moreover, at the approximate minimum of the $E_{tot}$ curve, the pressure
does not vanish.

The purpose of this paper is to analyze how these systematic
discontinuities and associated errors in the total energy and the pressure
can be dealt with.
In Sec.~\ref{sec:definitions}, we introduce the definitions
that will be used all along the paper,
including the basic concept of interpolation of energy.
In Sec.~\ref{sec:correx}, we present accurate, but CPU time-consuming,
formulas for corrections to energy and pressure.
In Sec. ~\ref{sec:corrie} and \ref{sec:corret},
approximations that make it simpler to use are examined.
Sec.~\ref{sec:corrie} is dedicated to a new technique based on a
scaling hypothesis ({S.H.}), allowing to compute the
energy curve for a unique reference volume,
whereas Sec.~\ref{sec:corret} treats, with the same notations, the corrections
proposed by Froyen and Cohen \cite{Froyen-Cohen86} to pressure
and by Francis and Payne \cite{Francis-Payne90} to energy.
We emphasize the theoretical differences between these techniques.
In Sec.~\ref{sec:applic}, we present the results obtained using the
scaling hypothesis for silicium, barium titanate, and helium.
In Sec.~\ref{sec:compare}, we compare the different techniques by
their effect on the calculation of $a_0$, $B_0$, and $B_{0}^{'}$
in bulk silicon.
In Sec.~\ref{sec:aniso}, we present a generalization
to anisotropic deformations.
We introduce a correction to the stress tensor and apply it in
the case of bulk silicon.
Finally, we present our conclusions in Sec.~\ref{sec:concl}.

\section{Definitions}
\label{sec:definitions}

Let $\overline{N}_{PW}^d \left({E_{cut}},{V} \right)$ be the mean number of
plane waves in the basis set (see Appendix \ref{app:meanonk} and
\cite{Note_npw})
for a given value of the cut-off energy ${E_{cut}}$ and of the volume $V$.
The subscript ``d'' stands for ``discontinuous''.
As the reciprocal lattice is discrete,
$\overline{N}_{PW}$ can stay constant for a range of cut-off energies,
or for different volumes.
On the other hand, it will change abruptly for some values of ${E_{cut}}$,
at fixed volume, or for some values of $V$, at fixed ${E_{cut}}$.
So, $\overline{N}_{PW}^d \left({E_{cut}},{V} \right)$ is a stair-like function
(see Fig.~\ref{fig:npwd}).

Let $\overline{N}_{PW}^c \left({E_{cut}},{V} \right)$ be the fictitious
number of plane waves in the basis set for a given value
of the cut-off energy ${E_{cut}}$
and of the volume $V$, if it were determined by the product of the density
of state in the reciprocal lattice by the volume of a sphere of
radius ${(2 E_{cut} )}^{1/2}$~:
\begin{equation}
\overline{N}_{PW}^c \left({E_{cut}},{V} \right) =
\left( \frac {V} {8 \pi ^3} \right)
\frac {4} {3} \pi \left( {(2 E_{cut} )}^{1/2} \right)^3=
\frac {V} {6 \pi ^2} {(2 E_{cut} )}^{3/2}
\label{eq:npwc}
\end{equation}
where the subscript ``c'' stands for ``continuous''.

Let $E_{cut}^c \left({\overline{N}_{PW}},{V} \right)$
and $E_{cut}^d \left({\overline{N}_{PW}},{V} \right)$
be the inverse of
$\overline{N}_{PW}^c \left({E_{cut}},{V} \right)$ and
$\overline{N}_{PW}^d \left({E_{cut}},{V} \right)$, respectively.
The first can trivially be obtained from Eq.~(\ref{eq:npwc})~:
\begin{equation}
E_{cut}^c \left({\overline{N}_{PW}},{V} \right) =
\frac {1} {2} {\left( \frac {6 \pi ^2 \overline{N}_{PW}} {V} \right) }^{2/3}
\label{eq:ecutc}
\end{equation}
whereas the second is defined for certain values of $\overline{N}_{PW}$
for a given volume $V$, and links these values to a semi-opened interval
$\left[ E_{cut}^1 , E_{cut}^2 \right[ $.

Let $E_{tot} \left[ {\overline{N}_{PW}},{V} \right]$
be the total energy that is calculated
when the number of plane waves used in an actual calculation, at volume
$V$, is ${\overline{N}_{PW}}$.

For a given volume, this function is defined only for
$\overline{N}_{PW}=\overline{N}_{PW}^d \left({E_{cut}},{V} \right)$.
In order to obtain a continuous curve, we interpolate through these points,
so that $E_{tot} \left[ {\overline{N}_{PW}},{V} \right]$
can be obtained at any value of $\overline{N}_{PW}$.
At the present stage of the discussion, the choice of the interpolation scheme
is not relevant (see Sec.~\ref{sec:applic}).
The introduction of such an interpolating energy curve was also performed
in other papers treating the problem
\cite{Froyen-Cohen86,Francis-Payne90}.

We also define the function
$E_{tot}^d \left\{ {E_{cut}},{V} \right\}$ which gives the total
energy obtained when the basis set is determined by $E_{cut}$
through Eq.~(\ref{eq:defpwset}).
This function is connected to
$E_{tot} \left[ {\overline{N}_{PW}},{V} \right]$ by~:
\begin{equation}
E_{tot}^d \left\{ {E_{cut}},{V} \right\} =
E_{tot} \left[ {\overline{N}_{PW}^d \left({E_{cut}},{V} \right)},{V} \right]
\,.
\label{eq:etotd}
\end{equation}

Since $\overline{N}_{PW}^d \left({E_{cut}},{V} \right)$
changes by abrupt jumps, this function is not the smooth
desired one, but a set of micro-curves which correspond to a fixed
$\overline{N}_{PW}$
(see Fig.~\ref{fig:segments}).

We finally introduce an ideal continuous function
$E_{tot}^c \left\{ {E_{cut}},{V} \right\}$
which corresponds to the energy of an hypothetical calculation,
at volume $V$, with $\overline{N}_{PW}^c \left({E_{cut}},{V} \right)$
plane waves.
It is connected to
$E_{tot} \left[ {\overline{N}_{PW}},{V} \right]$
by the following equation~:
\begin{equation}
E_{tot}^c \left\{ {E_{cut}},{V} \right\} =
E_{tot} \left[ {\overline{N}_{PW}^c \left({E_{cut}},{V} \right)},{V} \right]
\,.
\label{eq:etotc}
\end{equation}
The function $E_{tot}^c \left\{ {E_{cut}},{V} \right\}$
is the smooth function that should be used
for the determination of material properties,
whereas the one we get from an usual calculation is
$E_{tot}^d \left\{ {E_{cut}},{V} \right\}$.
Thus, the correction to energy aims at determining the function
$E_{tot}^c \left\{ {E_{cut}},{V} \right\}$ starting from
$E_{tot}^d \left\{ {E_{cut}},{V} \right\}$.

Using the same type of notations, we can now introduce the definitions
related to pressure.
Let $P \left[ {\overline{N}_{PW}},{V} \right]$ be the pressure
that is calculated by using the stress theorem
\cite{Nielsen-Martin83,Nielsen-Martin85a,Nielsen-Martin85b}
when the number of plane waves used in the calculation, at
volume $V$, is ${\overline{N}_{PW}}$.
This function is connected to
$E_{tot} \left[ {\overline{N}_{PW}},{V} \right]$ by~:
\begin{equation}
P \left[ {\overline{N}_{PW}^1},{V_1} \right] =-
\left(
 \frac {\partial {E_{tot} \left[ {\overline{N}_{PW}},{V} \right] }}
 {\partial {V}}
\right) _{\overline{N}_{PW}}
\hspace{0.4mm}
\left| \!
 \renewcommand{\arraystretch}{0.4}
 \begin{array}{l}
 {} \\ {} \\ {\scriptstyle V=V_1} \\
 {\scriptstyle \overline{N}_{PW}=\overline{N}_{PW}^1}
 \end{array}
 \renewcommand{\arraystretch}{1}
\right.
\label{eq:pressuren}
\end{equation}
where the subscript $\overline{N}_{PW}$
to the parentheses indicates that the derivative of
the total energy versus the volume is taken at constant number of plane
waves, and the subscripts ${V=V_1}$ and
${\overline{N}_{PW}=\overline{N}_{PW}^1}$ to the vertical
bar indicate where the previous expression is evaluated.
Let $P^d \left\{ {E_{cut}},{V} \right\}$ be the pressure that is calculated
with $E_{cut}$ and the volume $V$ as inputs.
This function is connected to $E_{tot}^d \left\{ {E_{cut}},{V} \right\}$ by~:
\begin{equation}
P^d \left\{ {E_{cut}^1},{V_1} \right\} =-
\left(
 \frac {\partial {E_{tot}^d \left\{ {E_{cut}},{V} \right\}}}
 {\partial {E_{cut}}}
\right) _{V}
\hspace{0.4mm}
\left| \!
 \renewcommand{\arraystretch}{0.4}
 \begin{array}{l}
 {} \\ {} \\ {\scriptstyle V=V_1} \\ {\scriptstyle E_{cut}=E_{cut}^1}
 \end{array}
 \renewcommand{\arraystretch}{1}
\right.
\,.
\end{equation}

In fact, this function is a set of micro-curves,
each of which corresponds to fixed $\overline{N}_{PW}$
(see Fig.~\ref{fig:segments}).
So let $P^c \left\{ {E_{cut}},{V} \right\} $ be the ideal continuous
pressure curve.
This function is connected to $E_{tot}^c \left\{ {E_{cut}},{V} \right\}$ by~:
\begin{equation}
P^c \left\{ {E_{cut}^1},{V_1} \right\} =-
\left(
 \frac {\partial {E_{tot}^c \left\{ {E_{cut}},{V} \right\} }}
 {\partial {E_{cut}}}
\right) _{V}
\hspace{0.4mm}
\left| \!
 \renewcommand{\arraystretch}{0.4}
 \begin{array}{l}
 {} \\ {} \\ {\scriptstyle V=V_1} \\ {\scriptstyle E_{cut}=E_{cut}^1}
 \end{array}
 \renewcommand{\arraystretch}{1}
\right.
\,.
\label{eq:pressurec}
\end{equation}
The function
$P^c \left\{ {E_{cut}},{V} \right\}$
is the function that should be used
for the determination of material properties,
whereas the one we get from an usual calculation is
$P^d \left\{ {E_{cut}},{V} \right\}$.
Thus, the correction to pressure aims at determining the function
$P^c \left\{ {E_{cut}},{V} \right\}$ starting from
$P^d \left\{ {E_{cut}},{V} \right\}$.

\section{Accurate Corrections to Energy and Pressure}
\label{sec:correx}

The correction to energy at $E_{cut}=E_{cut}^1$ and $V=V_1$,
needed to connect the output of a computer run
$E_{tot}^d \left\{ {E_{cut}^1},{V_1} \right\}$
with the ideal value $E_{tot}^c \left\{ {E_{cut}^1},{V_1} \right\} $
is obtained by combining Eqs.~(\ref{eq:etotd}) and (\ref{eq:etotc})~:
\begin{eqnarray}
E_{tot}^c \left\{ {E_{cut}^1},{V_1} \right\} & = &
E_{tot}^d \left\{ {E_{cut}^1},{V_1} \right\} \nonumber \\
& & + \left\{
E_{tot} \left[ {\overline{N}_{PW}^c \left({E_{cut}^1},{V_1} \right)},{V_1}
 \right] -
E_{tot} \left[ {\overline{N}_{PW}^d \left({E_{cut}^1},{V_1} \right)},{V_1}
 \right] \right\}
\,.
\label{eq:enepulcorbase}
\end{eqnarray}
A schematic representation of this expression is given in
Fig.~\ref{fig:enecor}.
Indeed, Eq.~(\ref{eq:enepulcorbase}) can be rewritten
$E_{tot}(B)=E_{tot}(C)+ \left\{ E_{tot}(D)-E_{tot}(E) \right\}$
following the notations of the figure.
The idea is that going from point C to B is equivalent to going from point
E to D, as suggested by the arrows in Fig.~\ref{fig:enecor}.

The correction to pressure (the correcting term is called Pulay stress
\cite{Francis-Payne90}
by analogy with the Pulay force \cite{Pulay69})
at $E_{cut}=E_{cut}^1$ and $V=V_1$
is obtained by deriving Eq.~(\ref{eq:enepulcorbase})
with respect to the volume at fixed cut-off energy~:
\begin{eqnarray}
P^c \left\{ {E_{cut}^1},{V_1} \right\} & = &
P^d \left\{ {E_{cut}^1},{V_1} \right\} \nonumber \\
& & +
P \left[ {\overline{N}_{PW}^c \left({E_{cut}^1},{V_1} \right)},{V_1} \right] -
P \left[ {\overline{N}_{PW}^d \left({E_{cut}^1},{V_1} \right)},{V_1} \right]
\nonumber \\
& & - \frac {\overline{N}_{PW}^c \left({E_{cut}^1},{V_1} \right)} {V_1}
\left(
 \frac {\partial {E_{tot} \left[ {\overline{N}_{PW}} {V} \right] }}
 {\partial {\overline{N}_{PW}}}
\right) _{V}
\hspace{0.4mm}
\left| \!
 \renewcommand{\arraystretch}{0.4}
 \begin{array}{l}
 {} \\ {} \\ {\scriptstyle V=V_1} \\
 {\scriptstyle
 \overline{N}_{PW}=\overline{N}_{PW}^c \left({E_{cut}^1},{V_1} \right)}
 \end{array}
 \renewcommand{\arraystretch}{1}
\right.
\,.
\label{eq:prepulcorbase}
\end{eqnarray}
This result was obtained by using the chain rule and the following
derivatives ~:
\begin{equation}
\left(
 \frac {\partial {\overline{N}_{PW}^c \left({E_{cut}},{V} \right)}}
 {\partial {V}}
\right) _{E_{cut}}=
\frac {\overline{N}_{PW}^c \left({E_{cut}},{V} \right)} {V}
\,,
\end{equation}
from Eq.~(\ref{eq:npwc}),
and
\begin{equation}
\left(
 \frac {\partial {\overline{N}_{PW}^d \left({E_{cut}},{V} \right)}}
 {\partial {V}}
\right) _{E_{cut}}=0
\,.
\end{equation}
since $\overline{N}_{PW}^d \left({E_{cut}},{V} \right)$ is a stair function.

The validity of Eqs.~(\ref{eq:enepulcorbase}) and (\ref{eq:prepulcorbase})
rests only on the choice of an interpolation scheme for
$E_{tot} \left[ {\overline{N}_{PW}},{V} \right] $.
At this stage, in order to find
$E_{tot}^c \left\{ {E_{cut}},{V} \right\}$ or
$P^c \left\{ {E_{cut}},{V} \right\}$ as function of the volume using
Eq.~(\ref{eq:enepulcorbase}) or Eq.~(\ref{eq:prepulcorbase}),
one should consider a few values of $V$, then for each $V$,
choose a few basis sets, and interpolate the energy to get
$E_{tot} \left[ {\overline{N}_{PW}},{V} \right]$.
Unfortunately, this procedure is time-consuming.

\section{Scaling hypothesis}
\label{sec:corrie}

We now introduce a technique that allows to make the interpolation effort
only for one given reference volume $V_0$.
It is different from the technique proposed by Francis and Payne for
correcting the energy \cite{Francis-Payne90}, or by Froyen and Cohen for
correcting the pressure \cite{Froyen-Cohen86} (see Sec.~\ref{sec:corret}).

We make a realistic hypothesis :
{\it for the purpose of the calculation of the correction
to energy and pressure},
the difference between energy at $V$
and at $V_0$ at constant $E_{cut}$ does not depend on $E_{cut}$~:
\begin{equation}
E_{tot}^c \left\{ {E_{cut}},{V} \right\} \approx
E_{tot}^c \left\{ {E_{cut}},{V_0} \right\} +
f(V-V_0) \hspace{15mm} \forall E_{cut}
\,.
\label{eq:hypbase}
\end{equation}
This is the mathematical expression of the principle of cancellation of
errors between similar geometries, already mentioned in the introduction.

By deriving this equation with respect to $E_{cut}$ at constant $V$, we get~:
\begin{equation}
\left(
 \frac {\partial {E_{tot}^c \left\{ {E_{cut}},{V} \right\} }}
 {\partial {E_{cut}}}
\right) _{V}
\hspace{0.4mm}
\left| \!
 \renewcommand{\arraystretch}{0.4}
 \begin{array}{l}
 {} \\ {} \\ {\scriptstyle V=V_1} \\ {\scriptstyle E_{cut}=E_{cut}^1}
 \end{array}
 \renewcommand{\arraystretch}{1}
\right.
\approx
\left(
 \frac {\partial {E_{tot}^c \left\{ {E_{cut}},{V} \right\} }}
 {\partial {E_{cut}}}
\right) _{V}
\hspace{0.4mm}
\left| \!
 \renewcommand{\arraystretch}{0.4}
 \begin{array}{l}
 {} \\ {} \\ {\scriptstyle V=V_0} \\ {\scriptstyle E_{cut}=E_{cut}^1}
 \end{array}
 \renewcommand{\arraystretch}{1}
\right.
\label{eq:hypecut}
\end{equation}
which means that the successive derivatives of $E_{tot}^c \left\{ {E_{cut}},{V}
\right\} $
with respect to the cut-off energy at constant volume do not depend on $V$
\cite{Note_hypecut}.
This last equation can be developped by means of Eq.~(\ref{eq:etotc}).
Deriving the latter with respect to $E_{cut}$ at constant $V$ and using
the chain rule, we get~:
\begin{eqnarray}
\left(
 \frac {\partial {E_{tot}^c \left\{ {E_{cut}},{V} \right\} }}
 {\partial {E_{cut}}}
\right) _{V}
& = &
\left(
 \frac {\partial {E_{tot} \left[ {\overline{N}_{PW}},{V} \right] }}
 {\partial {\overline{N}_{PW}}}
\right) _{V}
\hspace{0.4mm}
\left| \!
 \renewcommand{\arraystretch}{0.4}
 \begin{array}{l}
 {} \\ {} \\ {\scriptstyle V} \\
 {\scriptstyle
 \overline{N}_{PW}=\overline{N}_{PW}^c \left({E_{cut}},{V} \right)}
 \end{array}
 \renewcommand{\arraystretch}{1}
\right.
\left(
 \frac {\partial {\overline{N}_{PW}^c \left({E_{cut}},{V} \right)}}
 {\partial {E_{cut}}}
\right) _{V}
\nonumber \\
\end{eqnarray}
where the last term can be obtained from Eq.~(\ref{eq:npwc})~:
\begin{equation}
\left(
 \frac {\partial {\overline{N}_{PW}^c \left({E_{cut}},{V} \right)}}
 {\partial {E_{cut}}}
\right) _{V}=
\frac {V} {2 \pi ^2} {(2 E_{cut} )}^{1/2}
\,.
\end{equation}
When introducing these results in Eq.~(\ref{eq:hypecut}), we get~:
\begin{equation}
\left(
 \frac {\partial {E_{tot} \left[ {\overline{N}_{PW}},{V} \right] }}
 {\partial {\overline{N}_{PW}}}
\right) _{V}
\hspace{0.4mm}
\left| \!
 \renewcommand{\arraystretch}{0.4}
 \begin{array}{l}
 {} \\ {} \\ {\scriptstyle V=V_1} \\
 {\scriptstyle \overline{N}_{PW}=\overline{N}_{PW}^1}
 \end{array}
 \renewcommand{\arraystretch}{1}
\right.
\approx
\frac {V_0} {V_1}
\left(
 \frac {\partial {E_{tot} \left[ {\overline{N}_{PW}},{V} \right] }}
 {\partial {\overline{N}_{PW}}}
\right) _{V}
\hspace{0.4mm}
\left| \!
 \renewcommand{\arraystretch}{0.4}
 \begin{array}{l}
 {} \\ {} \\ {\scriptstyle V=V_0} \\
 {\scriptstyle \overline{N}_{PW}=\overline{N}_{PW}^0}
 \end{array}
 \renewcommand{\arraystretch}{1}
\right.
\label{eq:hypnpw}
\end{equation}
where
$\overline{N}_{PW}^1=\overline{N}_{PW}^c \left({E_{cut}^1},{V_1} \right)$
and
$\overline{N}_{PW}^0=\overline{N}_{PW}^c \left({E_{cut}^1},{V_0} \right)
=\frac {V_0} {V_1} \overline{N}_{PW}^1$.
The relation expressed in Eq.~(\ref{eq:hypnpw}) is valid for any
$\overline{N}_{PW}^1$ and $\overline{N}_{PW}^0$ connected by~:
\begin{equation}
\overline{N}_{PW}^0=\frac {V_0} {V_1} \overline{N}_{PW}^1
\,.
\label{eq:npwratio}
\end{equation}

Let us now consider the correction to energy.
We can rewrite Eq.~(\ref{eq:enepulcorbase}) in the following way~:
\begin{eqnarray}
E_{tot}^c \left\{ {E_{cut}},{V_1} \right\} & = &
E_{tot}^d \left\{ {E_{cut}},{V_1} \right\} \nonumber \\
& & +
\int _{\overline{N}_{PW}^d \left({E_{cut}},{V_1} \right)}
^{\overline{N}_{PW}^c \left({E_{cut}},{V_1} \right)}
\renewcommand{\arraystretch}{0.001}
\begin{array}{l} \\
{\left(
 \frac {\partial {E_{tot} \left[ {\overline{N}_{PW}},{V} \right] }}
 {\partial {\overline{N}_{PW}}}
\right) _{V}
 \hspace{0.4mm}
\left| \!
 \renewcommand{\arraystretch}{0.2}
 \begin{array}{l}
 {} \\ {} \\ {\scriptstyle V=V_1} \\
 {\scriptstyle \overline{N}_{PW}=\overline{N}_{PW}^1}
 \end{array}
 \renewcommand{\arraystretch}{1}
\right.}
\:{\scriptstyle d\,{\overline{N}_{PW}^1}}Ê\\
\\
\end{array}
\renewcommand{\arraystretch}{1}
\,.
\label{eq:enepulcorinteg}
\end{eqnarray}
Inserting Eq.~(\ref{eq:hypnpw}) in this expression, we get~:
\begin{eqnarray}
E_{tot}^c \left\{ {E_{cut}},{V_1} \right\} & \approx &
E_{tot}^d \left\{ {E_{cut}},{V_1} \right\} \nonumber \\
& & +
\int _{\overline{N}_{PW}^d \left({E_{cut}},{V_1} \right)}
^{\overline{N}_{PW}^c \left({E_{cut}},{V_1} \right)}
\renewcommand{\arraystretch}{0.001}
\begin{array}{l} \\
{\frac {V_0} {V_1}
 \left(
 \frac {\partial {E_{tot} \left[ {\overline{N}_{PW}} {V} \right] }}
 {\partial {\overline{N}_{PW}}}
\right) _{V}
\hspace{0.4mm}
\left| \!
 \renewcommand{\arraystretch}{0.2}
 \begin{array}{l}
 {} \\ {} \\ {\scriptstyle V=V_0} \\
 {\scriptstyle \overline{N}_{PW}=\frac {V_0} {V_1}\overline{N}_{PW}^1}
 \end{array}
 \renewcommand{\arraystretch}{1}
\right.}
\:{\scriptstyle d\,{\overline{N}_{PW}^1}}Ê\\
\\
\end{array}
\renewcommand{\arraystretch}{1}
\,.
\end{eqnarray}
By introducing the change of variables given by Eq.~(\ref{eq:npwratio}),
we get~:
\begin{eqnarray}
E_{tot}^c \left\{ {E_{cut}},{V_1} \right\} & \approx &
E_{tot}^d \left\{ {E_{cut}},{V_1} \right\} \nonumber \\
& & +
\int _{\frac {V_0} {V_1} \overline{N}_{PW}^d \left({E_{cut}},{V_1} \right)}
^{\frac {V_0} {V_1} \overline{N}_{PW}^c \left({E_{cut}},{V_1} \right)}
\renewcommand{\arraystretch}{0.001}
\begin{array}{l} \\
{\left(
 \frac {\partial {E_{tot} \left[ {\overline{N}_{PW}},{V} \right] }}
 {\partial {\overline{N}_{PW}}}
\right) _{V}
 \hspace{0.4mm}
\left| \!
 \renewcommand{\arraystretch}{0.2}
 \begin{array}{l}
 {} \\ {} \\ {\scriptstyle V=V_0} \\
 {\scriptstyle \overline{N}_{PW}=\overline{N}_{PW}^0}
 \end{array}
 \renewcommand{\arraystretch}{1}
\right.}
\:{\scriptstyle d\,{\overline{N}_{PW}^0}}Ê\\
\\
\end{array}
\renewcommand{\arraystretch}{1}
\,.
\end{eqnarray}
Finally, we have~:
\begin{eqnarray}
E_{tot}^c \left\{ {E_{cut}},{V_1} \right\} & \approx &
E_{tot}^d \left\{ {E_{cut}},{V_1} \right\} \nonumber \\
& & +
E_{tot} \left[ {\frac {V_0} {V_1}
                   \overline{N}_{PW}^c \left({E_{cut}},{V_1} \right)},{V_0}
         \right] -
E_{tot} \left[ {\frac {V_0} {V_1}
                   \overline{N}_{PW}^d \left({E_{cut}},{V_1} \right)},{V_0}
         \right]
\,.
\label{eq:enepulcorvref}
\end{eqnarray}
This expression allows to correct the value of energy obtained at $V_1$
by using the interpolating curve $E_{tot} \left[ {\overline{N}_{PW}},{V}
\right] $
calculated at $V_0$.

Let us finally consider the correction to pressure.
We can get different expressions depending on the way we work starting
from Eq.~(\ref{eq:enepulcorbase}).
We can either use the scaling approximation and then differentiate
with respect to the volume, or vice versa.
In the first case, we start from Eq.~(\ref{eq:enepulcorvref})
and differentiate it with respect to the volume
at fixed cut-off energy.
This leads to~:
\begin{eqnarray}
P^c \left\{ {E_{cut}^1},{V_1} \right\} & \approx &
P^d \left\{ {E_{cut}^1},{V_1} \right\} \nonumber \\
& & - \frac {V_0} {{V_1}^2} {\overline{N}_{PW}^d \left({E_{cut}^1},{V_1}
\right)}
\left(
 \frac {\partial {E_{tot} \left[ {\overline{N}_{PW}},{V} \right] }}
 {\partial {\overline{N}_{PW}}}
\right) _{V}
\hspace{0.4mm}
\left| \!
 \renewcommand{\arraystretch}{0.4}
 \begin{array}{l}
 {} \\ {} \\ {\scriptstyle V=V_0} \\
 {\scriptstyle
 \overline{N}_{PW}=
 \frac {V_0} {V_1} \overline{N}_{PW}^d \left({E_{cut}^1},{V_1} \right) }
 \end{array}
 \renewcommand{\arraystretch}{1}
\right.
\,.
\nonumber \\
\label{eq:prepulcorvref}
\end{eqnarray}
Using Eq.~(\ref{eq:hypnpw}), this can be rewritten~:
\begin{eqnarray}
P^c \left\{ {E_{cut}^1},{V_1} \right\} & \approx &
P^d \left\{ {E_{cut}^1},{V_1} \right\} \nonumber \\
& & - \frac {\overline{N}_{PW}^d \left({E_{cut}^1},{V_1} \right)} {V_1}
\left(
 \frac {\partial {E_{tot} \left[ {\overline{N}_{PW}},{V} \right] }}
 {\partial {\overline{N}_{PW}}}
\right) _{V}
\left| \!
 \renewcommand{\arraystretch}{0.4}
 \begin{array}{l}
 {} \\ {} \\ {\scriptstyle V=V_1} \\
 {\scriptstyle
 \overline{N}_{PW}=\overline{N}_{PW}^d \left({E_{cut}^1},{V_1} \right)}
 \end{array}
 \renewcommand{\arraystretch}{1}
\right.
\,.
\end{eqnarray}
On the other side, we can start from Eq.~(\ref{eq:prepulcorbase}) and
use the approximation Eq.~(\ref{eq:hypecut}) and its consequences.
This leads to~:
\begin{eqnarray}
P^c \left\{ {E_{cut}^1},{V_1} \right\} & \approx &
P^d \left\{ {E_{cut}^1},{V_1} \right\} \nonumber \\
& & + \frac {V_1} {V_0} \left(
P \left[ {\frac {V_0} {V_1}
            \overline{N}_{PW}^c \left({E_{cut}^1},{V_1} \right)},{V_0}
   \right] -
P \left[ {\frac {V_0} {V_1}
            \overline{N}_{PW}^d \left({E_{cut}^1},{V_1} \right)},{V_0}
   \right]
\right) \nonumber \\
& & - \frac {V_0} {{V_1}^2} {\overline{N}_{PW}^c \left({E_{cut}^1},{V_1}
\right)}
\left(
 \frac {\partial {E_{tot} \left[ {\overline{N}_{PW}},{V} \right] }}
 {\partial {\overline{N}_{PW}}}
\right) _{V}
\hspace{0.4mm}
\left| \!
 \renewcommand{\arraystretch}{0.4}
 \begin{array}{l}
 {} \\ {} \\ {\scriptstyle V=V_0} \\
 {\scriptstyle
 \overline{N}_{PW}=
 \frac {V_0} {V_1} \overline{N}_{PW}^c \left({E_{cut}^1},{V_1} \right)}
 \end{array}
 \renewcommand{\arraystretch}{1}
\right.
\,.
\nonumber \\
\label{eq:prepulcorvbis}
\end{eqnarray}
Between these different, but equivalent, expressions of the correction to
pressure
Eq.~(\ref{eq:prepulcorvref}) is the most convenient because it simply
needs the interpolation of $E_{tot}$ as a function of $\overline{N}_{PW}$
for a given reference volume $V_0$.

\section{Comparison with other Techniques}
\label{sec:corret}

While the scaling hypothesis ({S.H.}), presented in the preceding section,
starts from the correction to energy and proceeds by derivation to obtain
the correction to pressure,
Francis and Payne \cite{Francis-Payne90} have proposed a technique
for the correction to energy that starts from the correction to pressure
proposed by Froyen and Cohen \cite{Froyen-Cohen86}
and proceeds by integration.
Let us recall their results.

\subsection{Froyen-Cohen correction to pressure}

The expression proposed by Froyen and Cohen
for the correction to pressure is the following~:
\begin{eqnarray}
P^c_{FC} \left\{ {E_{cut}^1},{V_1} \right\} & = &
P^d \left\{ {E_{cut}^1},{V_1} \right\} \nonumber \\
& & - \frac {2} {3} \frac {E_{cut}^1} {V_1}
\left(
 \frac {\partial {E_{tot}^c \left\{ {E_{cut}},{V} \right\} }}
 {\partial {E_{cut}}}
\right) _{V}
\left| \!
 \renewcommand{\arraystretch}{0.4}
 \begin{array}{l}
 {} \\ {} \\ {\scriptstyle V=V_1} \\ {\scriptstyle E_{cut}=E_{cut}^1}
 \end{array}
 \renewcommand{\arraystretch}{1}
\right.
\,.
\label{eq:prepulcorfcecut}
\end{eqnarray}
The definition of an interpolating energy curve is also central in
this approach.

Let us develop Eq.~(\ref{eq:prepulcorfcecut}) to compare it with
the accurate correction to pressure Eq.~(\ref{eq:prepulcorbase})
and our approximations Eqs.~(\ref{eq:prepulcorvref})
and (\ref{eq:prepulcorvbis}).
By using the chain rule in Eq.~(\ref{eq:prepulcorfcecut}), we easily get~:
\begin{eqnarray}
P^c_{FC} \left\{ {E_{cut}^1},{V_1} \right\} & = &
P^d \left\{ {E_{cut}^1},{V_1} \right\} \nonumber \\
& & - \frac {\overline{N}_{PW}^c \left({E_{cut}^1},{V_1} \right)} {V_1}
\left(
 \frac {\partial {E_{tot} \left[ {\overline{N}_{PW}},{V} \right] }}
 {\partial {\overline{N}_{PW}}}
\right) _{V}
\hspace{0.4mm}
\left| \!
 \renewcommand{\arraystretch}{0.4}
 \begin{array}{l}
 {} \\ {} \\ {\scriptstyle V=V_1} \\
 {\scriptstyle
 \overline{N}_{PW}=\overline{N}_{PW}^c \left({E_{cut}^1},{V_1} \right)}
 \end{array}
 \renewcommand{\arraystretch}{1}
\right.
\label{eq:prepulcorfcnpw}
\end{eqnarray}
where one term is missing compared to Eq.~(\ref{eq:prepulcorbase}).

Let us continue the analysis.
By using the scaling hypothesis in the form of Eq.~(\ref{eq:hypecut})
and Eq.~(\ref{eq:hypnpw}) respectively
in Eq.~(\ref{eq:prepulcorfcecut}) and Eq.~(\ref{eq:prepulcorfcnpw}), we get~:
\begin{eqnarray}
P^c_{FC} \left\{ {E_{cut}^1},{V_1} \right\} & = &
P^d \left\{ {E_{cut}^1} {V_1} \right\} \nonumber \\
& & - \frac {2} {3} \frac {E_{cut}^1} {V_1}
\left(
 \frac {\partial {E_{tot}^c \left\{ {E_{cut}},{V} \right\} }}
 {\partial {E_{cut}}}
\right) _{V}
\hspace{0.4mm}
\left| \!
 \renewcommand{\arraystretch}{0.4}
 \begin{array}{l}
 {} \\ {} \\ {\scriptstyle V=V_0} \\ {\scriptstyle E_{cut}=E_{cut}^1}
 \end{array}
 \renewcommand{\arraystretch}{1}
\right.
, \\
& = &
P^d \left\{ {E_{cut}^1},{V_1} \right\} \nonumber \\
& & - \frac {V_0} {{V_1}^2}
{\overline{N}_{PW}^c \left({E_{cut}^1},{V_1} \right)}
\left(
 \frac {\partial {E_{tot} \left[ {\overline{N}_{PW}},{V} \right] }}
 {\partial {\overline{N}_{PW}}}
\right) _{V}
\hspace{0.4mm}
\left| \!
 \renewcommand{\arraystretch}{0.4}
 \begin{array}{l}
 {} \\ {} \\ {\scriptstyle V=V_0} \\
 {\scriptstyle
 \overline{N}_{PW}=
 \frac {V_0} {V_1} \overline{N}_{PW}^c \left({E_{cut}^1},{V_1} \right)}
 \end{array}
 \renewcommand{\arraystretch}{1}
\right.
\label{eq:prepulcorfcvref}
\end{eqnarray}
where clearly one term is missing in regard to Eq.~(\ref{eq:prepulcorvbis}).
When this equation is compared to Eq.~(\ref{eq:prepulcorvref}), we see that
$\overline{N}_{PW}^d \left({E_{cut}^1},{V_1} \right)$ is changed into
$\overline{N}_{PW}^c \left({E_{cut}^1},{V_1} \right)$.
This is due to an inaccurate definition of the Pulay stress
(as shown in Appendix \ref{app:froyen-cohen}).

Between these different expressions of the correction to pressure
Eq.~(\ref{eq:prepulcorfcvref}) is the most convenient because it just
needs an interpolation of $E_{tot}$ as a function of $\overline{N}_{PW}$
for a given reference volume $V_0$.

Before going further, we illustrate the two proposed stress correction
techniques in Fig.~\ref{fig:sipre}.
We draw the output of a computer run $P^d \left\{ {E_{cut}^1},{V_1} \right\} $
and
the curves $P^c \left\{ {E_{cut}^1},{V_1} \right\} $ obtained respectively by
applying
to it either the correction given by Eq.~(\ref{eq:prepulcorvref}) or
Froyen-Cohen correction given by Eq.~(\ref{eq:prepulcorfcvref}).
We also draw the curve obtained by applying the following pressure correction~:
\begin{eqnarray}
P^* \left\{ {E_{cut}^1},{V_1} \right\} & = &
P^d \left\{ {E_{cut}^1},{V_1} \right\} \nonumber \\
& & + \frac {V_1} {V_0} \left(
P \left[ {\frac {V_0} {V_1}
            \overline{N}_{PW}^c \left({E_{cut}^1},{V_1} \right)},{V_0}
   \right] -
P \left[ {\frac {V_0} {V_1}
            \overline{N}_{PW}^d \left({E_{cut}^1},{V_1} \right)},{V_0}
   \right]
\right)
\,.
\label{eq:prepulcorpar}
\end{eqnarray}
which consists in adding to $P^d \left\{ {E_{cut}^1},{V_1} \right\} $
the missing term in Eq.~(\ref{eq:prepulcorfcvref})
in regard to Eq.~(\ref{eq:prepulcorvbis})
(which is equivalent to Eq.~(\ref{eq:prepulcorvref})).
The graph clearly shows the importance of each term of the proposed
corrections.
The correction of Froyen and Cohen
given by Eq.~(\ref{eq:prepulcorfcvref}) is responsible of a
shift of the uncorrected curve.
Whereas, the correction given by Eq.~(\ref{eq:prepulcorpar}) is
responsible of the cancellation of the jumps between the micro-curves.
The {S.H.} correction includes these two effects as it
can be seen from Eq.~(\ref{eq:prepulcorvbis}) and Fig.~\ref{fig:sipre}.
The first is definitely the most important, but the second is not negligible.
Moreover, Eq.~(\ref{eq:prepulcorvref}) or Eq.~(\ref{eq:prepulcorfcvref})
are equally easy to use, so that the more accurate Eq.~(\ref{eq:prepulcorvref})
should always be preferred.

\subsection{Francis-Payne correction to energy}

The expression proposed by Francis and Payne for the correction to energy is
the following (see Appendix \ref{app:francis-payne}):
\begin{eqnarray}
E_{tot,FP}^c \left\{ {E_{cut}^1},{V_1} \right\} & = &
E_{tot}^d \left\{ {E_{cut}^1},{V_1} \right\} \nonumber \\
& & - \frac {2 E_{cut}^1} {3}
\ln \left( {\frac {\overline{N}_{PW}^d \left({E_{cut}^1},{V_1} \right)}
{\overline{N}_{PW}^c \left({E_{cut}^1},{V_1} \right)}} \right)
\left(
 \frac {\partial {E_{tot}^c \left\{ {E_{cut}} {V} \right\} }}
 {\partial {E_{cut}}}
\right) _{V}
\hspace{0.4mm}
\left| \!
 \renewcommand{\arraystretch}{0.4}
 \begin{array}{l}
 {} \\ {} \\ {\scriptstyle V=V_1} \\ {\scriptstyle E_{cut}=E_{cut}^1}
 \end{array}
 \renewcommand{\arraystretch}{1}
\right.
\,.
\nonumber \\
\label{eq:enepulcorfpecut}
\end{eqnarray}
It can be shown (see Appendix \ref{app:francis-payne}) that
the derivation of this expression implies the use of the approximation given
by Eq.~(\ref{eq:hypecut}).
Moreover, it is obtained by integrating the Pulay stress expression
proposed by Froyen and Cohen which is inaccurate
(see Appendix \ref{app:froyen-cohen}).
By using the chain rule in Eq.~(\ref{eq:enepulcorfpecut}), we get~:
\begin{eqnarray}
E_{tot,FP}^c \left\{ {E_{cut}^1},{V_1} \right\} &=&
E_{tot}^d \left\{ {E_{cut}^1} {V_1} \right\}
\nonumber \\
& & - \overline{N}_{PW}^c \left({E_{cut}^1},{V_1} \right)
\ln \left( {\frac {\overline{N}_{PW}^d \left({E_{cut}^1},{V_1} \right)}
{\overline{N}_{PW}^c \left({E_{cut}^1},{V_1} \right)}} \right)
\nonumber \\
& & \times
\left(
 \frac {\partial {E_{tot} \left[ {\overline{N}_{PW}},{V} \right] }}
 {\partial {\overline{N}_{PW}}}
\right) _{V}
\hspace{0.4mm}
\left| \!
 \renewcommand{\arraystretch}{0.4}
 \begin{array}{l}
 {} \\ {} \\ {\scriptstyle V=V_1} \\
 {\scriptstyle
 \overline{N}_{PW}=\overline{N}_{PW}^c \left({E_{cut}^1},{V_1} \right)}
 \end{array}
 \renewcommand{\arraystretch}{1}
\right.
\,.
\label{eq:enepulcorfpnpw}
\end{eqnarray}
By using our approximation in the form of Eq.~(\ref{eq:hypecut})
in Eq.~(\ref{eq:enepulcorfpecut}), we get~:
\begin{eqnarray}
E_{tot,FP}^c \left\{ {E_{cut}^1},{V_1} \right\} & = &
E_{tot}^d \left\{ {E_{cut}^1},{V_1} \right\} \nonumber \\
& & - \frac {2 E_{cut}^1} {3}
\ln \left( {\frac {\overline{N}_{PW}^d \left({E_{cut}^1},{V_1} \right)}
{\overline{N}_{PW}^c \left({E_{cut}^1},{V_1} \right)}} \right)
\left(
 \frac {\partial {E_{tot}^c \left\{ {E_{cut}},{V} \right\} }}
 {\partial {E_{cut}}}
\right) _{V}
\hspace{0.4mm}
\left| \!
 \renewcommand{\arraystretch}{0.4}
 \begin{array}{l}
 {} \\ {} \\ {\scriptstyle V=V_0} \\ {\scriptstyle E_{cut}=E_{cut}^1}
 \end{array}
 \renewcommand{\arraystretch}{1}
\right.
\,.
\nonumber \\
\end{eqnarray}
By using our approximation in the form of Eq.~(\ref{eq:hypnpw})
in Eq.~(\ref{eq:enepulcorfpnpw}), we get~:
\begin{eqnarray}
E_{tot,FP}^c \left\{ {E_{cut}^1},{V_1} \right\} &=&
E_{tot}^d \left\{ {E_{cut}^1},{V_1} \right\}
\nonumber \\
& & - \frac {V_0} {V_1}
\overline{N}_{PW}^c \left({E_{cut}^1},{V_1} \right)
\ln \left( {\frac {\overline{N}_{PW}^d \left({E_{cut}^1},{V_1} \right)}
{\overline{N}_{PW}^c \left({E_{cut}^1},{V_1} \right)}} \right)
\nonumber \\
& & \times
\left(
 \frac {\partial {E_{tot} \left[ {\overline{N}_{PW}},{V} \right] }}
 {\partial {\overline{N}_{PW}}}
\right) _{V}
\hspace{0.4mm}
\left| \!
 \renewcommand{\arraystretch}{0.4}
 \begin{array}{l}
 {} \\ {} \\ {\scriptstyle V=V_0} \\
 {\scriptstyle
 \overline{N}_{PW}=
 \frac {V_0} {V_1} \overline{N}_{PW}^c \left({E_{cut}^1},{V_1} \right)}
 \end{array}
 \renewcommand{\arraystretch}{1}
\right.
\,.
\label{eq:enepulcorfpvref}
\end{eqnarray}
This last expression of the correction to energy is the most convenient
because it just needs an interpolation of $E_{tot}$ as a function of
$\overline{N}_{PW}$ for a given reference volume $V_0$.

The comparison with {S.H.} technique is not as straightforward as for
correction to pressure.
But it is clear that the inaccuracy in the definition of the Pulay
stress, from which Eq.~(\ref{eq:enepulcorfpvref}) is derived,
is a source of error.

\section{Applications}
\label{sec:applic}

The importance of the correction is now investigated for silicon,
barium titanate, and helium, various materials presenting different
types of bonding.

Our calculations, performed
within the local density approximation (LDA) \cite{Kohn-Sham65}, use a
preconditioned conjugate gradient algorithm \cite{Teter89,Payne92}.
We use a rational polynomial parametrization of the
exchange-correlation energy functional \cite{Teter}, which is based
on the Ceperley-Alder gas data \cite{Ceperley-Alder80}.
The Brillouin zone is sampled with different Monkhorst-Pack
\cite{Monkhorst-Pack76} meshes of special $k$-points.
The ``all-electron'' potentials are replaced by {\it ab initio}, separable,
norm-conserving pseudopotentials, as described below.

Then we apply Eqs.~(\ref{eq:enepulcorvref}) and (\ref{eq:prepulcorvref})
to correct the results.
The interpolating scheme for obtaining
$E_{tot} \left[ {\overline{N}_{PW}},{V_0} \right] $ is the following.
We first calculate the total energy for three different values of
$\overline{N}_{PW}$~:
\begin{equation}
\overline{N}_{PW}^d \left({E_{cut}-3\%},{V_0} \right),
\overline{N}_{PW}^d \left({E_{cut}},{V_0} \right),
\overline{N}_{PW}^d \left({E_{cut}+3\%},{V_0} \right)
\,.
\nonumber
\end{equation}
Then we interpolate between these values by an exponential fit of the form~:
\begin{equation}
E_{tot} \left[ {\overline{N}_{PW}},{V_0} \right] =
E_{tot}^{inf}+\exp \left( a_0+a_1 \overline{N}_{PW} \right)
\,.
\end{equation}

First, we apply this correction to bulk silicon, with covalent bonding.
The corresponding pseudopotential was built following the scheme proposed
in Ref.~\cite{Hamann89}.
The atomic positions in the unit cell are completely
determined by symmetry~:
the only free parameter is the lattice parameter $a$.
We compute simultaneously the total energy $E_{tot}(a)$ and the
pressure $P(a)$ as we vary the lattice parameter $a$ at a constant energy
cut-off ($E_{cut}$=6 Ha with 2 special $k$-points in the irreducible
Brillouin zone).
Then we apply {S.H.} correction.
The effect of correction to energy, in Fig.~\ref{fig:siene} is obvious~:
it cancels out the jumps between micro-curves.
The effect of correction to pressure in Fig.~\ref{fig:sipre}
has already been commented.
Firstly, the jumps between micro-curves are suppressed.
Secondly, the whole curve is shifted towards larger pressure.
These results will be commented with more details in Sec.~\ref{sec:compare}.
The lattice parameter value is found to be 10.23 Bohr.

Then, we consider barium titanate (BaTiO$_{3}$), for which we used
extended norm-conserving pseudopotentials given in Ref.~\cite{Ghosez94}.
This material is sometimes classified as an ionic compound.
As pointed out recently, it also presents a partial covalent character
\cite{Cohen92,Ghosez95}.
Its structure is cubic perovskite at high temperature
(above about 120$^\circ$C) while it undergoes 3 successive ferroelectric
phase transitions as the temperature goes down.
In this study, we will focus on the cubic, high symmetry phase in which
the atoms are at symmetric positions so that
the only structural degree of freedom is the lattice parameter $a$.
When using the {S.H.} correction, a first estimation of $a_0$ at 7.53 Bohr
is already possible on the basis of only 4 points at 20 Ha cutoff with 4
points in the irreducible Brillouin zone (see Fig.~\ref{fig:tiene20}).
For BaTiO$_{3}$, an accurate investigation of some properties (as, for example,
the phonons frequencies) requires nevertheless to work at
a 45 hartrees energy cut-off on a $6\times 6 \times 6$ mesh of $k$-points.
At this high cut-off, the total energy is well converged
($\frac {\partial E_{tot}}{\partial \overline{N}_{PW}}$ is very small) so that
the correction on $E_{tot}$ becomes negligible (Fig.~\ref{fig:tiful}).
We predict a lattice parameter of 7.45 Bohr in good agreement
with other LDA calculations.
By contrast, even for this case, the error on the pressure remains critical.
As illustrated on Fig. \ref{fig:tiful}, it can be efficiently corrected
when using the {S.H.} technique.
This correct estimation of the pressure
reveals essential for the investigation of
the energy surface of the low symmetry phases.

Finally, we perform an energetic calculation for FCC helium.
Once again, we observe the cancellation of the scattering
in the total energy data and the shift of the pressure curve,
showing that the correction is independent of the type of bonding, and
applicable to a large range of materials.
These results will be published in a future paper
\cite{Note_helium}.

\section{Comparison of the techniques}
\label{sec:compare}

It has been shown in the previous section that corrections to energy
and stresses can be rather important.
Now, we analyze in more detail their importance, and we compare
more quantitatively the two possible correction techniques
in the case of silicon
(Eqs.~(\ref{eq:enepulcorvref}) and (\ref{eq:prepulcorvref})
for the scaling hypothesis technique,
Eqs.~(\ref{eq:prepulcorfcvref}) and (\ref{eq:enepulcorfpvref})
for Froyen-Cohen technique and Francis-Payne technique).
We investigate different aspects of this correction,
more specifically we analyze its effect on the calculation of
the lattice constant $a_0$, the bulk modulus $B_0$, and its
pressure derivative $B_{0}^{'}$.
These properties can easily be determined either
from the curve of energy $E_{tot}$ versus lattice constant $a$
(or, in an equivalent way, versus volume as $V \propto {a^3}$)
or from that of pressure $P$ versus lattice constant $a$.

In order to obtain the $E_{tot} (a)$ and $P (a)$ curves,
we calculate the total energy and the pressure for a set of lattice constants
at a given cut-off energy.

As we mentioned above, the result is a set
of micro-curves~:
$E_{tot}^d \left\{ {E_{cut}},{a} \right\} $ and
$P^d \left\{ {E_{cut}},{a} \right\} $.
By fitting a polynomial to these values, we get continuous
curves $E_{tot} (a)$ and $P (a)$.
Depending on its degree, the polynomial fit matches more or
less accurately the data.
We measure the matching by the standard deviation $\chi$ of
the data $\left(x_i,y_i\right)$ from the polynomial $g(x)$~:
\begin{equation}
\chi=\sqrt{\frac
 {\renewcommand{\arraystretch}{0.2}
 \begin{array}{c}
 {\scriptstyle N} \\
 \Sigma \\
 {\scriptstyle i=1} \\
 \end{array}
 \renewcommand{\arraystretch}{1}
 \left({y_i-g(x_i)}\right)^2}
 {N-1}}
\label{eq:sdev}
\end{equation}
where $N$ is the number of data.
For example, from Fig.~\ref{fig:siene}, we have calculated the value of $\chi$
for respectively uncorrected data (open circles) and those obtained with {S.H.}
correction (solid circles), with respect to their corresponding polynomial fit
(respectively, the broken curve and the solid one).
We illustrate in Fig.~\ref{fig:noise} the evolution of $\chi$ as a
function of the degree of the polynomial.
On one side, it is evident that the higher the degree of the polynomial is,
the better the matching with the data will be.
On the other side, the aim of the fitting is to get rid off of the
noise \cite{Note_noise} due to jumps between micro-curves.
The higher the degree, the greater the part of this noise included
in the fit.
So, a compromise has to be reached.
We will consider that the
separation between trustable data and the noise has been accomplished
as soon as the standard deviation reaches a plateau (called residual noise)
as a function of the degree of the polynomial.
Going to higher-order polynomial would mean beginning to include the
residual noise in the fit.
So, the degree of the polynomial is chosen by detection of a plateau in the
standard deviation of the data from the polynomial.
In Fig.~\ref{fig:noise}, this plateau is already reached for a polynomial
of the third degree.

We apply the different techniques presented in the previous sections
to correct the calculated values.
By fitting a polynomial to these values, we get two different expressions of
$E_{tot}^c \left\{ {E_{cut}},{a} \right\} $ and of
$P^c \left\{ {E_{cut}},{a} \right\} $, corresponding to the
Froyen-Cohen-Francis-Payne correction or to the scaling hypothesis.
So, finally, we have three different expressions of the $E_{tot} (a)$ and
$P (a)$ curves.

However, if there are not enough data, statistical fluctuations occur
on the static equilibrium properties ($a_0$, $B_0$, and $B_{0}^{'}$)
due to the noise \cite{Note_noise} present when the correction is not used
(see Figs.~\ref{fig:siene}-\ref{fig:tiful} of Sec.~\ref{sec:applic}
especially for BaTiO$_3$).
So, from now on, we will work with a large number of data points in order
to suppress these fluctuations, and concentrate on the techniques.

\subsection{Scale Analysis}

In order to analyze quantitatively the reduction of noise,
we consider different sets of lattice parameter,
each set being characterized by a given scale which is defined
as the distance between two successive points of the set.
We consider six different scales~: 0.01, 0.025, 0.05, 0.1, 0.2, and 0.5 Bohr
(see Table~\ref{tab:scales}).
We generate the three different $E_{tot} (a)$ curves
and of the three $P (a)$ ones for all these sets.
{}From these curves, we determine
the lattice constant $a_0$, the bulk modulus $B_0$
and its pressure derivative $B_{0}^{'}$.

Many observations occur.
The residual noise in the energy and pressure data
is decreased by the correction.
This noise reduction is more important for small scales than large ones
(see Fig.~\ref{fig:noisered}).
This is due to the fact that the presence of micro-curves is much more
apparent at small than at large scales.
So, for the {\it energy} curve, the results obtained at large scales with and
without correction do not differ significantly.
Whereas, for the {\it pressure} curve, the correction also includes a shift
of the uncorrected curve~:
whatever the scale the correction is always important.
So, from now on, we just discuss the results obtained with the corrected values
of
energy and pressure.
Regarding the lattice constant, it can be seen (see Table~\ref{tab:ascales})
that there is a quite good agreement between the values obtained at the
different scales and between those obtained starting
from the $E_{tot} (a)$ curve and those calculated from the $P (a)$ one.
Regarding the bulk modulus, there is still a very good agreement
between the values obtained at the different scales from the pressure curve
but not between those calculated from the energy one
(see Table~\ref{tab:bscales}).
Moreover, the values obtained starting from the $E_{tot} (a)$ curve and those
calculated from the $P (a)$ curve agree only at large scale.

Regarding the derivative of the bulk modulus versus pressure,
the agreement is worse (these data are not reproduced here).

As a general rule for the calculation of static
equilibrium properties ($a_0$, $B_0$, and $B_{0}^{'}$),
we can state that
firstly, calculations based on the pressure data are more accurate than those
based on energy data,
and secondly, the accuracy decreases with the number of
derivatives taken from the starting curve.

Regarding the different proposed techniques, it can be noticed that their
results are nearly the same.
This is due to the fitting operation that eliminates the residual noise.
Note, nevertheless, that it is always lower with the {S.H.} technique,
especially for the pressure curve (see Fig.~\ref{fig:sipre}).

\subsection{Micro-curve Analysis}
When studying the small scales, in the previous section, micro-curves
corresponding to a constant number of plane waves have been detected.
The analysis of these micro-curves separately will emphasize
the efficiency of the different corrections.

For this purpose, we consider the scale 0.01 Bohr with the following
sets of points for each micro-curve
(see open circles in Fig.~\ref{fig:siene})~:
from 10.00 to 10.04, from 10.05 to 10.12, from 10.13 to 10.21,
from 10.22 to 10.29, from 10.30 to 10.37, from 10.38 to 10.44,
and from 10.45 to 10.50.
We generate for each of these micro-curves the three different
$E_{tot} (a)$ curves and of the three $P (a)$ ones.
{}From these curves, we also determine
the lattice constant $a_0$,
the bulk modulus $B_0$, and its pressure derivative $B_{0}^{'}$.

Regarding the lattice constant, the agreement between the values
obtained for the different micro-curves (see mean value $\mu$ and
standard deviation $\sigma$ in Table~\ref{tab:asegments})
is much better with {S.H.} technique than
with Froyen-Cohen or Francis-Payne technique.
Note also that there is a very good consistency between the values obtained
starting from the $E_{tot} (a)$ curve and those calculated from the $P (a)$
one.

Regarding the bulk modulus, there is still a very good accordance
between the values obtained for the different micro-curves
from the pressure curve
but it is a little bit worse for those calculated from the energy one
(see mean value $\mu$ and standard deviation $\sigma$
in Table~\ref{tab:bsegments}).
Moreover, the values obtained starting from the $E_{tot} (a)$ curve and those
calculated from the $P (a)$ one agree very well.

Regarding the derivative of the bulk modulus versus pressure, the
agreement is worse, except for results obtained by calculations based on
the pressure curve with the {S.H.} technique
(these data are not reproduced here).

The conclusions, regarding the calculation of static
equilibrium properties ($a_0$, $B_0$, and $B_{0}^{'}$),
drawn in the scale analysis still apply for micro-curve analysis.

Regarding the different techniques, the results are generally not the same.
This is due to the fact these techniques transform the
original micro-curve into different micro-curves.
The agreement of the results between the micro-curves is a mesure of the
effectiveness of the correction method.
In this case, it clearly shows that the {S.H.} is the best.
When looking at Fig.~\ref{fig:siene}, we see that there are still
discontinuities (this should be analysed by further studies)
in $E_{tot} (a)$ curve though there is a good agreement of the slopes between
the micro-curves.
This means that the micro-curves are parallel, which is confirmed by the fact
that there are no discontinuities in the $P (a)$ curve.

We have also tested the validity of the approximation given by
Eq.~(\ref{eq:hypecut}).
The first test consists in using a reference volume different for each
micro-curve.
Doing so, the difference between $V$ and $V_0$ is smaller
and thus the influence of the approximation should be reduced.
It appears that the results obtained (not reproduced here) are almost not
affected, in favor of the scaling hypothesis Eq.~(\ref{eq:hypecut}).
The second test consists in not introducing the approximation.
This can only be done for a reduced set of lattice constants, for example
a micro-curve, because it necessitates to calculate
$E_{tot} \left[ {\overline{N}_{PW}},{V} \right] $
for each of the lattice constants.
Unfortunately, this increases the residual noise so that no comparison
can be made, and this second test is inconclusive.

\subsection{Convergence Analysis}
We finally analyze the effect of the correction on the convergence of
the calculated values of $a_0$, $B_0$, and $B_0^{'}$ in cut-off energy
(working successively at 3, 6, 10, and 15 Ha, the latter considered as giving
completely converged values) and number of special $k$-points
(working successively with 2, 6, and 10 special points,
the latter considered as giving completely converged values) for three
different scales (0.01, 0.1, and 0.5).
We see that working at $E_{cut}$=10 Ha with 2 special $k$-points
can be considered as enough converged (it leads to a relative error of
0.1\% on $a_0$, 1\% on $B_0$, and $B_{0}^{'}$).
{}From Tables~\ref{tab:aconverge} and~\ref{tab:bconverge}, we see that the
convergence is not really improved by the correction except for the
result obtained from the pressure curve.
This can be explained by the fact that the fitting operation acts as a
correction by eliminating the residual noise on the energy curve, so that
the results with and without correction are nearly the same.
In the limit of a very large number of data points, this implicit
correction is sufficient.
However, in practical applications, very few points are used,
and the correction is needed to avoid the statistical fluctuations
mentioned above.
In this sense, it can be said to improve the convergence.
For the pressure, the correction also includes a shift of the
uncorrected curve, this explains why the correction is always important
whatever the degree of convergence.

\section{Anisotropic deformations}
\label{sec:aniso}
We now consider anisotropic deformations and generalize our technique for
stress correction.
In this case, the total energy does not only depend on the volume of the
unit cell but also on its shape.
In order to be as general as possible, we consider that $E_{tot}$ depends
on the matrix $\underline{\underline {A}}$ formed by the components of
unit cell vectors.
So the energy correction given by Eq.~(\ref{eq:enepulcorbase}) becomes in the
case of isotropic deformations~:
\begin{eqnarray}
E_{tot}^c \left\{ {E_{cut}^1},{\underline{\underline {A}}^1} \right\} & = &
E_{tot}^d \left\{ {E_{cut}^1},{\underline{\underline {A}}^1} \right\}
\nonumber \\
& & +
E_{tot} \left[{\overline{N}_{PW}^c \left({E_{cut}^1},{V_1} \right)},
                  {\underline{\underline {A}}^1}
         \right] -
E_{tot} \left[{\overline{N}_{PW}^d \left({E_{cut}^1},{V_1} \right)},
                  {\underline{\underline {A}}^1}
         \right]
\,.
\label{eq:enepulcorbasegene}
\end{eqnarray}
where $V_1=\det \underline{\underline {A}}^1$ is the volume of the unit cell.
We now make the approximation that the difference between energy at
$\underline{\underline {A}}$
and at $\underline{\underline {A}}^0$
at constant $E_{cut}$ does not depend on $E_{cut}$~:
\begin{equation}
E_{tot}^c \left\{ {E_{cut}},{\underline{\underline {A}}} \right\} \approx
E_{tot}^c \left\{ {E_{cut}},{\underline{\underline {A}}^0} \right\} +
f(\underline{\underline {A}} \rightarrow\underline{\underline {A}}^0)
\hspace{15mm} \forall E_{cut}
\end{equation}
which is a generalization of Eq.~(\ref{eq:hypbase}).
This leads to the generalized forms of Eq.~(\ref{eq:hypecut})~:
\begin{equation}
\left(
 \frac {\partial {E_{tot}^c \left\{ {E_{cut}},{\underline{\underline {A}}}
 \right\} }}
 {\partial {E_{cut}}}
\right) _{\underline{\underline {A}}}
\hspace{0.4mm}
\left| \!
 \renewcommand{\arraystretch}{0.4}
 \begin{array}{l}
 {} \\ {} \\ {\scriptstyle \underline{\underline {A}}=
 \underline{\underline {A}}^1} \\
 {\scriptstyle E_{cut}=E_{cut}^1}
 \end{array}
 \renewcommand{\arraystretch}{1}
\right.
\approx
\left(
 \frac {\partial {E_{tot}^c \left\{ {E_{cut}},
 {\underline{\underline {A}}} \right\} }}
 {\partial {E_{cut}}}
\right) _{\underline{\underline {A}}}
\hspace{0.4mm}
\left| \!
 \renewcommand{\arraystretch}{0.4}
 \begin{array}{l}
 {} \\ {} \\ {\scriptstyle \underline{\underline {A}}=
 \underline{\underline {A}}^0} \\
 {\scriptstyle E_{cut}=E_{cut}^1}
 \end{array}
 \renewcommand{\arraystretch}{1}
\right.
\end{equation}
 and of Eq.~(\ref{eq:hypnpw})~:
\begin{equation}
\left(
 \frac {\partial {E_{tot} \left[ {\overline{N}_{PW}},
 {\underline{\underline {A}}} \right] }}
 {\partial {\overline{N}_{PW}}}
\right) _{\underline{\underline {A}}}
\hspace{0.4mm}
\left| \!
 \renewcommand{\arraystretch}{0.4}
 \begin{array}{l}
 {} \\ {} \\ {\scriptstyle \underline{\underline {A}}=
 \underline{\underline {A}}^1} \\
 {\scriptstyle \overline{N}_{PW}=\overline{N}_{PW}^1}
 \end{array}
 \renewcommand{\arraystretch}{1}
\right.
\approx
\frac {V_0} {V_1}
\left(
 \frac {\partial {E_{tot} \left[ {\overline{N}_{PW}},
 {\underline{\underline {A}}} \right] }}
 {\partial {\overline{N}_{PW}}}
\right) _{\underline{\underline {A}}}
\hspace{0.4mm}
\left| \!
 \renewcommand{\arraystretch}{0.4}
 \begin{array}{l}
 {} \\ {} \\ {\scriptstyle \underline{\underline {A}}=
 \underline{\underline {A}}^0} \\
 {\scriptstyle \overline{N}_{PW}=\overline{N}_{PW}^0}
 \end{array}
 \renewcommand{\arraystretch}{1}
\right.
\end{equation}
where
\begin{equation}
\overline{N}_{PW}^0=\frac {V_0} {V_1} \overline{N}_{PW}^1
 =\frac {\det \underline{\underline {A}}^0}
 {\det \underline{\underline {A}}^1} \overline{N}_{PW}^1
\,.
\end{equation}

Finally, the generalized energy correction writes~:
\begin{eqnarray}
E_{tot}^c \left\{ {E_{cut}},{\underline{\underline {A}}^1} \right\} & \approx &
E_{tot}^d \left\{ {E_{cut}},{\underline{\underline {A}}^1} \right\} \nonumber
\\
& & +
E_{tot} \left[ {\frac {V_0} {V_1}
                  \overline{N}_{PW}^c \left({E_{cut}},{V_1} \right)},
                  {\underline{\underline {A}}^0}
         \right] -
E_{tot} \left[ {\frac {V_0} {V_1}
                   \overline{N}_{PW}^d \left({E_{cut}},{V_1} \right)},
                   {\underline{\underline {A}}^0}
         \right]
\,.
\end{eqnarray}

Deriving this expression with respect to the strain
$\epsilon_{\alpha \beta}$ at fixed
cut-off energy, we get the stress correction~:
\begin{eqnarray}
\lefteqn{
\sigma_{\alpha \beta}^c \left\{ {E_{cut}^1},
{\underline{\underline {A}}^1} \right\} \approx
\sigma_{\alpha \beta}^d \left\{ {E_{cut}^1},
{\underline{\underline {A}}^1} \right\} }
\hspace{15mm}
\nonumber \\
& & + \delta_{\alpha \beta}
\frac {V_0} {{V_1}^2} {\overline{N}_{PW}^d \left({E_{cut}^1},{V_1} \right)}
\left(
 \frac {\partial {E_{tot} \left[ {\overline{N}_{PW}},
 {\underline{\underline {A}}} \right] }}
 {\partial {\overline{N}_{PW}}}
\right) _{\underline{\underline {A}}}
\hspace{0.4mm}
\left| \!
 \renewcommand{\arraystretch}{0.4}
 \begin{array}{l}
 {} \\ {} \\ {\scriptstyle \underline{\underline {A}}=
 \underline{\underline {A}}^0} \\
 {\scriptstyle
 \overline{N}_{PW}=\frac {V_0} {V_1}
 \overline{N}_{PW}^d \left({E_{cut}^1},{V_1} \right)}
 \end{array}
 \renewcommand{\arraystretch}{1}
\right.
\label{eq:stresscor}
\end{eqnarray}
where we have used the following definitions~:
\begin{equation}
\sigma_{\alpha \beta}^d \left\{ {E_{cut}^1},
{\underline{\underline {A}}^1} \right\} =
\left(
 \frac {\partial {E_{tot}^d \left\{ {E_{cut}},
 {\underline{\underline {A}}} \right\} }}
 {\partial {\epsilon_{\alpha \beta}}}
\right) _{E_{cut}}
\hspace{0.4mm}
\left| \!
 \renewcommand{\arraystretch}{0.4}
 \begin{array}{l}
 {} \\ {\scriptstyle \underline{\underline {\epsilon}}=0} \\
 {\scriptstyle \underline{\underline {A}}=\underline{\underline {A}}^1} \\
 {\scriptstyle E_{cut}=E_{cut}^1}
 \end{array}
 \renewcommand{\arraystretch}{1}
\right.
\,,
\end{equation}
and
\begin{equation}
\sigma_{\alpha \beta}^c \left\{ {E_{cut}^1},
{\underline{\underline {A}}^1} \right\} =
\left(
 \frac {\partial {E_{tot}^c \left\{ {E_{cut}},
 {\underline{\underline {A}}} \right\} }}
 {\partial {\epsilon_{\alpha \beta}}}
\right) _{E_{cut}}
\hspace{0.4mm}
\left| \!
 \renewcommand{\arraystretch}{0.4}
 \begin{array}{l}
 {} \\ {\scriptstyle \underline{\underline {\epsilon}}=0} \\
 {\scriptstyle \underline{\underline {A}}=\underline{\underline {A}}^1} \\
 {\scriptstyle E_{cut}=E_{cut}^1}
 \end{array}
 \renewcommand{\arraystretch}{1}
\right.
\,.
\end{equation}
We have also used the following result~:
\begin{equation}
\frac {1} {V_1}
\frac {\partial V} {\partial \epsilon_{\alpha \beta}}
\hspace{0.4mm}
\left| \!
 \renewcommand{\arraystretch}{0.35}
 \begin{array}{l}
 {} \\ {} \\ {\scriptstyle \underline{\underline {\epsilon}}=0} \\
 {\scriptstyle V=V_1}
 \end{array}
 \renewcommand{\arraystretch}{1}
\right.
=\delta_{\alpha \beta}
\,.
\end{equation}
which is obtained by deriving the expression of
the volume $V$ of a unit cell defined by matrix $\underline{\underline {A}}$
in terms of the strain tensor
$\underline{\underline {\epsilon}}$ with respect to
the unit cell defined by matrix $\underline{\underline {A}}^1$~:
\begin{equation}
V=\det \underline{\underline {A}}
 =V_1 \det (\underline{\underline {\delta}}+\underline{\underline {\epsilon}})
 =\frac {1} {6} V_1 \varepsilon_{ijk} \varepsilon_{lmn}
 (\delta_{il}+\epsilon_{il})
 (\delta_{jm}+\epsilon_{jm})
 (\delta_{kn}+\epsilon_{kn})
 \,.
\end{equation}
The third-rank tensor $\underline{\underline{\underline {\varepsilon}}}$
is defined in order that $\varepsilon_{ijk}$ is equal
to +1 if \{i,j,k\} corresponds to any even permutation of \{1,2,3\},
to -1 if \{i,j,k\} corresponds to any odd permutation of \{1,2,3\},
to 0 in every other case.

It can be shown from Eq.~(\ref{eq:stresscor}) that the stress correction
only concerns the diagonal part of the stress tensor $\sigma_{\alpha \beta}$.
This result had already been mentioned by Vanderbilt \cite{Vanderbilt87},
but in a form similar to that proposed by Froyen and Cohen
\cite{Froyen-Cohen86}.
More precisely, the stress tensor $\sigma_{\alpha \beta}$ can be decomposed
into an isotropic contribution $\sigma_{\alpha \beta}^h$
(where the h superscript stands for ``hydrostatic'')
and an anisotropic one $\sigma_{\alpha \beta}^{d}$
(where the d superscript stands for ``deviatory'')~:
\begin{equation}
\sigma_{\alpha \beta}^h=-P \delta_{\alpha \beta}
\end{equation}
 and
\begin{equation}
\sigma_{\alpha \beta}^d=\sigma_{\alpha \beta}-\sigma_{\alpha \beta}^h
\end{equation}
where P is the pressure defined by Eq.~(\ref{eq:pressdef}).
{}From this decomposition, it appears that only the isotropic part
$\sigma_{\alpha \beta}^h$ of the stress tensor has to be corrected.
So, we come back to the case of isotropic deformations and the
correction to pressure is given by Eq.~(\ref{eq:prepulcorvref}).

We apply this correction in the case of silicon.
The chosen anisotropic deformation corresponds to a compression
(or expansion) of
the unit cell along the $[0 0 1]$ direction, with concurrent
expansion (or compression) of the unit cell along the
$[1 0 0]$ and $[0 1 0]$ directions,
such that the length of the three cubic directions are changed
to $a=b$ and $c$, while the volume $V$ is unchanged.
The energy and the pressure are functions of the ratio $y=b/c$.
$\sigma_{11}=\sigma_{22}$ and $\sigma_{33}$ stresses are present, while
non-diagonal stresses vanish.
Note that the atomic positions in the unit cell are no longer completely
determined by symmetry.
The results obtained are shown in Fig.~\ref{fig:sianiso}.
The minimum of the total energy and the zero of stresses nearly correspond to
the expected value of $y=1$ (the relative error is $0.3\%$ for the value
obtained from the energy curve and $0.03\%$ for the values obtained from
stresses curves).
The effect of energy correction in Fig.~\ref{fig:sianiso} is obvious.
In the stress graph, first, there is a suppression of
jumps between micro-curves.
Though it is not visible on the graph, it can be detected by the residual
noise reduction (it is approximatively divided by 3).
Second, there is a shift of the uncorrected curve.
It is the only effect included in Froyen-Cohen technique.
In this particular case, it is a shift by a constant due to the fact that
the deformation is at constant volume.
Indeed, $\overline{N}_{PW}^c \left({E_{cut}},{V} \right)$ is constant
in Eq.~(\ref{eq:prepulcorfcvref}) for Froyen-Cohen technique and
in Eq.~(\ref{eq:prepulcorvref}) for {S.H.} technique.

\section{Conclusions}
\label{sec:concl}

In this paper, a new means of correcting total energy and stresses
calculations performed using a fixed cut-off energy plane wave basis
set has been presented.
This technique relies on the interpolation of the energy as a function of the
number of plane wave, and a scaling hypothesis (S.H.) that allows to work with
a unique reference volume.
It has been compared to that of Froyen and Cohen for
stress correction and that of Francis and Payne for energy correction,
both presented using the same notations.
On a theoretical point of view, the approximations used in the different
approaches have been contrasted, showing that the scaling hypothesis
technique is more rigorous.
On a practical point of view, we have shown the importance of the correction
by presenting its effects on different materials (Si, BaTiO$_3$, and He)
corresponding to different types of bonding.

Then, we have compared the different methods by analyzing their effect on the
calculation of $a_0$, $B_0$, and $B_{0}^{'}$ in bulk silicium starting
from both the energy curve and the pressure one.
As a general rule, calculations based on the pressure data are more
accurate than those based on the energy data.
Also, the accuracy reduces with the number of derivatives
taken from the starting curve (i.e. pressure derivative of the bulk
modulus is more difficult to
correct than the bulk modulus).
The results of a scale analysis shows that if we do not focus on one
micro-curve, the benefits of the different correction techniques
are rather similiar.
This due to the final fitting operation that cancels the residual noise,
although the latter is always obtained lower with the {S.H.} technique.
By contrast, the results obtained when working at the micro-curve level show
that the {S.H.} is the best, though there are still discontinuities in
$E_{tot} (a)$ curve.

Regarding the convergence with respect to the cut-off energy and to the
number of special $k$-points, we see that it is not really improved
by the corrections except for the results obtained by the pressure curve.
For the energy curve, in the limit of a very large number of data points,
the fitting operation acts as a correction by eliminating the residual noise,
and the results with and without correction are nearly the same.
However, when working with a small number of data, the correction
is needed to avoid statistical fluctuations.
For the stress curve, the correction also includes a shift of the
uncorrected curve, this explains why the correction is always important
whatever the degree of convergence.

\acknowledgments
We thank J.-M. Beuken for permanent computer assistance.
We are grateful to D. C. Allan for useful discussions.
Our computational results were obtained using a version of the software program
Plane\_Wave (written by D. C. Allan),
which is marketed by Biosym Technologies of San Diego.
Three of the authors (G.-M. R., J.-C. C. and X. G.) have benefited from
financial support of the National Fund for Scientific Research (Belgium).
This paper presents research results of Belgian Program on
Interuniversity Attraction Poles initiated by the Belgian State, Prime
Minister's Office, Science Policy Programming.
We also acknowledge the use of the RS 6000 workstation from the common
project between IBM Belgium, UCL and FUNDP.


\appendix

\section{Mean number of plane waves}
\label{app:meanonk}
Since the number of plane waves differs at each $k$-point,
while the techniques exposed in this paper use a single number to
characterize this set,
we decide to work with an average number of plane waves.
It is not clear how this average should be calculated
on the different $k$-points.
We can think to an arithmetic mean~:
\begin{equation}
\overline{N}_{PW}^d \left({E_{cut}},{V} \right)=
\sum_{k} f_k
N_{PW,k}^d \left( E_{cut},V \right).
\end{equation}
or to a geometric mean (as suggested by Francis and Payne
\cite{Francis-Payne90})~:
\begin{equation}
\overline{N}_{PW}^d \left({E_{cut}},{V} \right) =
\prod_{k}
\left( N_{PW,k}^d \left( E_{cut},V \right) \right)^{f_k}.
\end{equation}
where $f_k$ are weights defined in order that~:
\begin{equation}
\sum_{k} f_k = 1
\,.
\end{equation}.

An interesting starting point for our reflexion would be the expression
of the total energy as function of its value at the different $k$-points.
Such an expression is known for some of its components.
For example, the kinetic energy is the arithmetic mean
of its value $E_{kin,k}$ at the different $k$-points~:
\begin{equation}
E_{kin} =
\sum_{k} f_k
E_{kin,k}
\label{eq:ekinmean}
\end{equation}
where $f_k$ are the weights of the different $k$-points.
On the other hand, the exchange-correlation contribution to total energy
is global;
it can not be physically divided in a sum of contributions of each $k$-point.
However, we can suppose that Eq.~(\ref{eq:ekinmean}) can be generalized
for total energy, so that we can write~:
\begin{equation}
E_{tot}^d \left\{ {E_{cut}},{V} \right\} =
\sum_{k} f_k
E_{tot,k}^d \left\{ E_{cut},V \right\},
\label{eq:etotmean}
\end{equation}
By using Eq.~(\ref{eq:etotd}) to develop Eq.~(\ref{eq:etotmean}), we get~:
\begin{equation}
E_{tot} \left[ {\overline{N}_{PW}^d \left({E_{cut}},{V} \right)},{V} \right] =
\sum_{k} f_k
E_{tot,k} \left[ {N_{PW,k}^d \left( E_{cut},V \right)},{V} \right].
\label{eq:npwmean}
\end{equation}
This expression can be considered as definition of the average number of
plane waves.

Unfortunately, it is not very useful as $E_{tot,k}$ is not known
as a function of $N_{PW}$ at each different $k$-point.
To say something about the definition of $\overline{N}_{PW}$, we have to
consider cases where the total energy at each different $k$-point and the
total energy $E_{tot}$ are the same function of $N_{PW}$
(it corresponds to deleting the $k$ subscript of $E_{tot}$ in the
right-hand side of Eq.~(\ref{eq:npwmean})).
This precisely the case where the bands are dispersionless, which is the case
e.g. when studying an isolated molecule in a supercell.
Let us analyze two of these simple cases.
If all these functions were the same linear function of $N_{PW}$,
the definition of $\overline{N}_{PW}$ given by
Eq.~(\ref{eq:npwmean}) would correspond
to an arithmetic mean.
Whereas, if all these functions were the same logarithmic function of $N_{PW}$,
the definition of $\overline{N}_{PW}$ given by Eq.~(\ref{eq:npwmean})
would correspond to a geometric mean.

In our calculations, we decide to use the geometric mean for defining
$\overline{N}_{PW}$.
This choice seems more physical than the arithmetic mean, because the
latter suppose a linear dependence of $E_{tot}$ versus $N_{PW}$.
But it is clear that the problem should be investigated further, and that
it is a source of error on the corrected values.

\section{Froyen-Cohen Technique for pressure correction}
\label{app:froyen-cohen}

The starting point of this technique is the definition of the Pulay Stress.
It is the quantity to be added to the pressure that is obtained
using $E_{cut}$ and $V$ for the definition of the plane-wave basis set,
to get the pressure that effectively corresponds
to these given values of $E_{cut}$ and $V$~:
\begin{eqnarray}
\sigma_{Pulay} \left( E_{cut}^1,V_1 \right) &=&
P^c \left\{ {E_{cut}^1},{V_1} \right\} -
P^d \left\{ {E_{cut}^1},{V_1} \right\}
\nonumber \\
&=&
P^c \left\{ {E_{cut}^1},{V_1} \right\} -
P \left[ {\overline{N}_{PW}^d \left({E_{cut}^1},{V_1} \right)},{V_1} \right] .
\label{eq:goodpul}
\end{eqnarray}
Froyen and Cohen, in their paper \cite{Froyen-Cohen86}, propose :
\begin{equation}
\sigma_{Pulay,FC} \left( E_{cut}^1,V_1 \right) =
- \frac {2} {3} \frac {E_{cut}^1} {V_1}
\left(
 \frac {\partial {E_{tot}^c \left\{ {E_{cut}},{V} \right\} }}
 {\partial {E_{cut}}}
\right) _{V}
\hspace{0.4mm}
\left| \!
 \renewcommand{\arraystretch}{0.4}
 \begin{array}{l}
 {} \\ {} \\ {\scriptstyle V=V_1} \\ {\scriptstyle E_{cut}=E_{cut}^1}
 \end{array}
 \renewcommand{\arraystretch}{1}
\right.
\,.
\label{eq:pulstress}
\end{equation}
We now show that this expression corresponds to an inaccurate
definition the Pulay stress.
Using the definition of $E_{cut}^c \left({\overline{N}_{PW}},{V} \right)$
given by Eq.~(\ref{eq:ecutc}), we get that~:
\begin{equation}
\left(
 \frac {\partial {E_{cut}^c \left({\overline{N}_{PW}},{V} \right)}}
 {\partial {V}}
\right) _{\overline{N}_{PW}}=
- \frac {1} {3} \left( 6 \pi^2 \overline{N}_{PW} \right)^{2/3}
\left( V \right)^{-5/3}=
- \frac {2} {3} \frac {E_{cut}^c \left({\overline{N}_{PW}},{V} \right)} {V}.
\end{equation}
The Pulay stress, as given by Eq.(~\ref{eq:pulstress})
can thus be rewritten as~:
\begin{eqnarray}
\sigma_{Pulay,FC} \left( E_{cut}^1,V_1 \right) &=&
\left(
 \frac {\partial {E_{tot}^c \left\{ {E_{cut}},{V} \right\} }}
 {\partial {E_{cut}}}
\right) _{V}
\hspace{0.4mm}
\left| \!
 \renewcommand{\arraystretch}{0.4}
 \begin{array}{l}
 {} \\ {} \\ {\scriptstyle V=V_1} \\ {\scriptstyle E_{cut}=E_{cut}^1}
 \end{array}
 \renewcommand{\arraystretch}{1}
\right.
\nonumber \\
& &\times
\left(
 \frac {\partial {E_{cut}^c \left({\overline{N}_{PW}},{V} \right)}}
 {\partial {V}}
\right) _{\overline{N}_{PW}}
\left| \!
 \renewcommand{\arraystretch}{0.4}
 \begin{array}{l}
 {} \\ {} \\ {\scriptstyle V=V_1} \\
 {\scriptstyle
 \overline{N}_{PW}=\overline{N}_{PW}^c \left({E_{cut}^1},{V_1} \right)}
 \end{array}
 \renewcommand{\arraystretch}{1}
\right.
\,.
\label{eq:pulstressfcb}
\end{eqnarray}
Finally, the expression of the Pulay stress given
by Eq.~(\ref{eq:pulstressfcb})
can be worked out, using the chain rule and the definitions
Eq.~(\ref{eq:pressuren}) and (\ref{eq:pressurec}) of
$P \left[ {\overline{N}_{PW}},{V} \right] $ and
$P^c \left\{ {E_{cut}},{V} \right\} $,
in order to give~:
\begin{equation}
\sigma_{Pulay,FC} \left( E_{cut}^1,V_1 \right) =
P^c \left\{ {E_{cut}^1},{V_1} \right\} -
P \left[ {\overline{N}_{PW}^c \left({E_{cut}^1},{V_1} \right)},{V_1} \right] .
\label{eq:pulstressfc}
\end{equation}
This expression does not correspond to the accurate definition of the
Pulay stress given by Eq.~(\ref{eq:goodpul}), since
$\overline{N}_{PW}^d \left({E_{cut}},{V} \right)$
has been replaced by $\overline{N}_{PW}^c \left({E_{cut}},{V} \right)$
in the last term of the right side of this expression.

\section{Francis-Payne Technique for energy correction}
\label{app:francis-payne}

The technique proposed by Francis and Payne consists in integrating
the Pulay stress in the form of Eq.~(\ref{eq:pulstressfc}) and
Eq.~(\ref{eq:pulstress}) between
$V_2$ and $V_1$ defined on Fig.~\ref{fig:enecor},
i.e. by the following relations~:
let $V_1$ be the volume at which energy has to be calculated,
$\overline{N}_{PW}^1=\overline{N}_{PW}^c \left({E_{cut}^1},{V_1} \right)$,
$\overline{N}_{PW}^2=\overline{N}_{PW}^d \left({E_{cut}^1},{V_1} \right)$ and
$V_2$ be the volume such that
$\overline{N}_{PW}^2=\overline{N}_{PW}^c \left({E_{cut}^1},{V_2} \right)$.

Let us now consider each terms of this integration.
The first term of the right side of
Eq.~(\ref{eq:pulstressfc}) is quite easy to integrate~:
\begin{eqnarray}
\int _{V_2}^{V_1}
\renewcommand{\arraystretch}{0.001}
\begin{array}{l} \\
{P^c \left\{ {E_{cut}^1},{V'} \right\} }
\:{\scriptstyle d\,{V'}}Ê\\
\\
\end{array}
\renewcommand{\arraystretch}{1}
& = &
\int _{V_2}^{V_1}
\renewcommand{\arraystretch}{0.001}
\begin{array}{l} \\
{-\left(
\frac {\partial {E_{tot}^c \left\{ {E_{cut}},{V} \right\} }}
 {\partial {V}}
 \right) _{E_{cut}}
 \left| \!
 \renewcommand{\arraystretch}{0.2}
 \begin{array}{l}
 {} \\ {} \\ {\scriptstyle V=V'} \\ {\scriptstyle E_{cut}=E_{cut}^1}
 \end{array}
 \renewcommand{\arraystretch}{1}
 \right.}
\:{\scriptstyle d\,{V'}}Ê\\
\\
\end{array}
\renewcommand{\arraystretch}{1}
\nonumber \\
& = &
E_{tot}^c \left\{ {E_{cut}^1},{V_2} \right\} -
E_{tot}^c \left\{ {E_{cut}^1},{V_1} \right\} \nonumber \\
& = &
E_{tot} \left[ {\overline{N}_{PW}^2},{V_2} \right] -
E_{tot} \left[ {\overline{N}_{PW}^1},{V_1} \right] .
\end{eqnarray}
It corresponds to $E_{tot}(A) - E_{tot}(B)$ on Fig.~\ref{fig:enecor}.

The integration of the right side of Eq.~(\ref{eq:pulstress})
necessitates the use of the approximation given by Eq.~(\ref{eq:hypecut})~:
\begin{eqnarray}
\lefteqn{
\int _{V_2}^{V_1}
\renewcommand{\arraystretch}{0.001}
\begin{array}{l} \\
 {- \frac {2} {3} \frac {E_{cut}^1} {V'}
 \left(
 \frac {\partial {E_{tot}^c \left\{ {E_{cut}},{V} \right\} }}
 {\partial {V}}
 \right) _{E_{cut}}
 \hspace{0.4mm}
 \left| \!
 \renewcommand{\arraystretch}{0.2}
 \begin{array}{l}
 {} \\ {} \\ {\scriptstyle V=V'} \\ {\scriptstyle E_{cut}=E_{cut}^1}
 \end{array}
 \renewcommand{\arraystretch}{1}
 \right.}
\:{\scriptstyle d\,{V'} =}Ê\\
\\
\end{array}
\renewcommand{\arraystretch}{1}
}
\hspace{35mm}
\nonumber \\
& &
\frac {2E_{cut}^1} {3} \ln \left( {\frac {V_2} {V_1}} \right)
\left(
 \frac {\partial {E_{tot}^c \left\{ {E_{cut}},{V} \right\} }}
 {\partial {E_{cut}}}
 \right) _{V}
\hspace{0.4mm}
\left| \!
 \renewcommand{\arraystretch}{0.4}
 \begin{array}{l}
 {} \\ {} \\ {\scriptstyle V=V_1} \\ {\scriptstyle E_{cut}=E_{cut}^1}
 \end{array}
 \renewcommand{\arraystretch}{1}
\right.
\end{eqnarray}
Using the definition of $V_1$ and $V_2$ with Eq.~(\ref{eq:npwc}), we get~:
\begin{equation}
\ln \left( {\frac {V_2} {V_1}} \right) =
\ln \left( {\frac {\overline{N}_{PW}^c \left({E_{cut}^1},{V_2} \right)}
{\overline{N}_{PW}^c \left({E_{cut}^1},{V_1} \right)}} \right) =
\ln \left( {\frac {\overline{N}_{PW}^d \left({E_{cut}^1},{V_1} \right)}
{\overline{N}_{PW}^c \left({E_{cut}^1},{V_1} \right)}} \right) =
\ln \left( {\frac {\overline{N}_{PW}^2} {\overline{N}_{PW}^1}} \right).
\end{equation}

A new approximation is needed to integrate of the second term
of the right side of Eq.~(\ref{eq:pulstressfc}), because it is impossible
to solve analytically~:
\begin{equation}
\int _{V_2}^{V_1}
\renewcommand{\arraystretch}{0.001}
\begin{array}{l} \\
{P \left[ {\overline{N}_{PW}},{V'} \right] }
\:{\scriptstyle d\,{V'}}Ê\\
\\
\end{array}
\renewcommand{\arraystretch}{1}
=
\int _{V_2}^{V_1}
\renewcommand{\arraystretch}{0.001}
\begin{array}{l} \\
{- \left(
 \frac {\partial {E_{tot} \left[ {\overline{N}_{PW}},{V} \right] }}
 {\partial {V}}
 \right) _{\overline{N}_{PW}}
 \hspace{0.4mm}
 \left| \!
 \renewcommand{\arraystretch}{0.2}
 \begin{array}{l}
 {} \\ {} \\ {\scriptstyle V=V'} \\
 {\scriptstyle
 \overline{N}_{PW}=\overline{N}_{PW}^c \left({E_{cut}^1},{V'} \right)}
 \end{array}
 \renewcommand{\arraystretch}{1}
\right.}
\:{\scriptstyle d\,{V'}}Ê\\
\\
\end{array}
\,.
\end{equation}
Indeed, the derivative in this integral has to be taken on a curve at
constant number of plane waves
$\overline{N}_{PW}=\overline{N}_{PW}^c \left({E_{cut}^1},{V'} \right)$,
but the latter varies while integrating (because of the variation of $V'$.
Graphically, it means that this derivative has to be evaluated
on each of the curves at constant $\overline{N}_{PW}$ situated between
the points $A$ and $B$ on Fig.~\ref{fig:enecor}.

On the other side, if we accept that the number of plane waves is
kept constant to
$\overline{N}_{PW}=\overline{N}_{PW}^c \left({E_{cut}^1},{V_2} \right)=
\overline{N}_{PW}^d \left({E_{cut}^1},{V_1} \right)=\overline{N}_{PW}^2$
while integrating, we get quite easily~:
\begin{equation}
\int _{V_2}^{V_1}
\renewcommand{\arraystretch}{0.001}
\begin{array}{l} \\
{P \left[ {\overline{N}_{PW}},{V'} \right] }
\:{\scriptstyle d\,{V'}}Ê\\
\\
\end{array}
\renewcommand{\arraystretch}{1}
=
E_{tot} \left[ {\overline{N}_{PW}^2},{V_2} \right] -
E_{tot} \left[ {\overline{N}_{PW}^2},{V_1} \right] .
\end{equation}
It corresponds to $E_{tot}(A) - E_{tot}(C)$ on Fig.~\ref{fig:enecor}.
It should be noted that making this approximation leads to a result
equivalent to the one that would be obtained by integrating
the Pulay stress in the form of Eq.~(\ref{eq:goodpul})
(instead of Eq.~(\ref{eq:pulstressfc})) and Eq.~(\ref{eq:pulstress}).

Globally, the integration of the right side of Eq.~(\ref{eq:pulstress})
gives $E_{tot}(B) - E_{tot}(C)$, which is the correction for energy~:
\begin{eqnarray}
E_{tot} \left[ {\overline{N}_{PW}^1},{V_1} \right] & = &
E_{tot} \left[ {\overline{N}_{PW}^2},{V_1} \right] \nonumber \\
& & - \frac {2 E_{cut}^1} {3}
\ln \left( {\frac {\overline{N}_{PW}^2} {\overline{N}_{PW}^1}} \right)
\left(
 \frac {\partial {E_{tot}^c \left\{ {E_{cut}},{V} \right\} }}
 {\partial {E_{cut}}}
\right) _{V}
\hspace{0.4mm}
 \left| \!
 \renewcommand{\arraystretch}{0.4}
 \begin{array}{l}
 {} \\ {} \\ {\scriptstyle V=V_1} \\ {\scriptstyle E_{cut}=E_{cut}^1}
 \end{array}
 \renewcommand{\arraystretch}{1}
 \right.
\,.
\end{eqnarray}
This can be rewritten as~:
\begin{eqnarray}
E_{tot}^c \left\{ {E_{cut}^1},{V_1} \right\} & = &
E_{tot}^d \left\{ {E_{cut}^1},{V_1} \right\} \nonumber \\
& & - \frac {2 E_{cut}^1} {3}
\ln \left( {\frac {\overline{N}_{PW}^d \left({E_{cut}^1},{V_1} \right)}
{\overline{N}_{PW}^c \left({E_{cut}^1},{V_1} \right)}} \right)
\left(
 \frac {\partial {E_{tot}^c \left\{ {E_{cut}},{V} \right\} }}
 {\partial {E_{cut}}}
\right) _{V}
\hspace{0.4mm}
 \left| \!
 \renewcommand{\arraystretch}{0.4}
 \begin{array}{l}
 {} \\ {} \\ {\scriptstyle V=V_1} \\ {\scriptstyle E_{cut}=E_{cut}^1}
 \end{array}
 \renewcommand{\arraystretch}{1}
 \right.
\end{eqnarray}
which is the final result presented by Francis and Payne.

\begin{figure}
\caption{Number of plane waves $\overline{N}_{PW}^{d}$ in a basis set defined
by Eq.~(\protect \ref{eq:defpwset}) in a 2D case.
The upper graph illustrates how these k-points are enclosed in a circle
of radius $(2 E_{cut})^{1/2}$.
The lower graph illustrates the stair-like evolution of
$\overline{N}_{PW}^{d}$ as a function of $E_{cut}$.}
\label{fig:npwd}
\end{figure}

\begin{figure}
\caption{Total energy (top) and pressure (bottom)
of Si calculated at constant cut-off energy
($E_{cut}$=6 Ha with 2 special $k$-points in the irreducible Brillouin zone
[IBZ]).
The graph highlights the presence of ``micro-curves'' (see text).}
\label{fig:segments}
\end{figure}

\begin{figure}
\caption{Schematic representation of energy correction.
The desired curve $E_{tot}^c \left\{ {E_{cut}^1},{V} \right\} $
is the one going through points A and B.
The calculated curve $E_{tot}^d \left\{ {E_{cut}^1},{V} \right\} $
is the one going through points A and C.
The correction at volume $V_1$ moves point C,
obtained by a calculation with
$\overline{N}_{PW}^2=\overline{N}_{PW}^d \left({E_{cut}^1},{V_1} \right)$,
to point B, obtained by an hypothetic calculation
with $\overline{N}_{PW}^1=\overline{N}_{PW}^c \left({E_{cut}^1},{V_1} \right)$.
This operation is equivalent to moving point E to point D.
This is easily done if the curve
$E_{tot} \left[ {\overline{N}_{PW}},{V_1} \right] $ is known
(see Eq.~(\protect \ref{eq:enepulcorbase}) ).}
\label{fig:enecor}
\end{figure}

\begin{figure}
\caption{Pressure in Si calculated at constant cut-off energy
($E_{cut}$=6 Ha with 2 special $k$-points in the IBZ) for 51 lattice constants.
(a) The open circles ($\circ$) represent the uncorrected values of pressure,
whereas the solid diamonds ($\blacklozenge$) illustrate the values corrected
by the Froyen-Cohen technique.
(b) The open diamonds ($\lozenge$) represent partially corrected values of
pressure
(see $P^*$ defined by Eq.~(\protect \ref{eq:prepulcorpar}) in the text),
whereas the solid circles ($\bullet$) are the values corrected by the
{S.H.} technique.
The graphs (a) and (b) point out that the {S.H.} correction
technique ($\bullet$) has two effects.
The first is a shift of the uncorrected curve ($\circ$), also
included in Froyen-Cohen technique ($\blacklozenge$).
Whereas the second is the cancellation of micro-curve jumps present
in the partial correction ($\lozenge$).}
\label{fig:sipre}
\end{figure}

\begin{figure}
\caption{Total energy of Si calculated at constant cut-off energy
($E_{cut}$=6 Ha with 2 special $k$-points in the IBZ) for 51 lattice constants.
The open circles ($\circ$) represent the uncorrected values of total energy,
whereas the solid circles ($\bullet$) illustrate the values corrected by the
{S.H.} technique.
The solid curve and (resp.) the broken curve are obtained by a
least-squares third-order fit to the corrected and (resp.) uncorrected data.
The solid curve yields a lattice parameter of 10.23 Bohr, whereas the
broken curve yields 10.17 Bohr.}
\label{fig:siene}
\end{figure}

\begin{figure}
\caption{Total energy of cubic BaTiO$_3$ calculated at constant cut-off energy
($E_{cut}$=20 Ha with 4 special $k$-points in the IBZ) for 4 lattice constants.
The open circles ($\circ$) represent the uncorrected values of total energy,
whereas the solid circles ($\bullet$) are the values corrected by the
{S.H.} technique.
The corrected values of energy are joined by a least-squares
third-order fit (solid curve).
Without correction, the residual noise leads to incorrect values
of the static equilibrium properties, as there are not enough data.
The lattice parameter value is found to be 7.53 Bohr from the solid line.}
\label{fig:tiene20}
\end{figure}

\begin{figure}
\caption{Total energy and pressure of cubic BaTiO$_3$ calculated at constant
cut-off energy
($E_{cut}$=45 Ha with 10 special $k$-points in the IBZ) for 8 lattice
constants.
The open circles ($\circ$) represent the uncorrected values
(joined by a scattered dashed line),
whereas the solid circles ($\bullet$) illustrate the values corrected by the
{S.H.} technique.
The corrected values of energy and pressure are joined
by a least-squares third-order fit(solid curves).
For total energy, the values obtained with or without
correction are nearly the same
(a high energy cut-off has been used, which leads
to well converged values of total energy).
For pressure, instead, the values obtained with or without
correction differ despite the high energy cut-off.
The lattice parameter value is found to be 7.45 Bohr from the solid lines.}
\label{fig:tiful}
\end{figure}

\begin{figure}
\caption{Semi-logarithmic plot of standard deviation $\chi$
of the data as a function of the degree of the polynomial
(obtained by least-squares fit).
The standard deviation $\chi$,
which is defined by Eq.~(\protect \ref{eq:sdev}) in the text,
is a measure of the matching of a polynomial to
the total energies of Si calculated at constant cut-off energy
($E_{cut}$=6 Ha with 2 special $k$-points in the IBZ) for 51 lattice constants
see Fig.~\protect \ref{fig:siene}.
The open circles ($\circ$) represent $\chi$
calculated from uncorrected values of total energy,
the open triangle ($\vartriangle$) represent that calculated from
the values corrected by Froyen-Cohen technique,
whereas the solid circles ($\bullet$) represent that calculated from
the values corrected by the {S.H.} technique.
The graph clearly illustrates the reduction of $\chi$ with the correction.}
\label{fig:noise}
\end{figure}

\begin{figure}
\caption{Plot of standard deviations $\chi_E$ (top) and $\chi_P$ (bottom)
of the energy and pressure data calculated at different scales
at constant cut-off energy
($E_{cut}$=6 Ha with 2 special $k$-points in the IBZ) for 51 lattice constants.
The white bars represent the standard deviation
$\chi$ calculated from uncorrected data,
the gray bars illustrate the one calculated from
the values corrected by Froyen-Cohen technique for pressure
and by Francis-Payne technique for energy,
whereas the black bars represent that calculated from
the values corrected by the {S.H.} technique.
The graph illustrates that the reduction of $\chi$ is more important for
small scales than large ones.
Note also that Froyen-Cohen technique does not reduce the noise in the pressure
data.}
\label{fig:noisered}
\end{figure}

\begin{figure}
\caption{Total energy (top) and $\sigma_{11}$ (middle) of Si
calculated at constant cut-off energy
($E_{cut}$=6 Ha with 2 special $k$-points in the IBZ) for 17 values of the
ratio $y=b/c$ in the case of an anistropic deformation which consists
in compressing (or expanding) the unit cell along the $[0 0 1]$ direction
in order that the length of the three cubic directions are changed
to $a=b$ and $c$ keeping the volume $V$ unchanged.
The open circles ($\circ$) represent the uncorrected values,
whereas the solid circles ($\bullet$) are the values corrected by the
{S.H.} technique.
The solid curves are obtained by least-squares third-order fit.
The graph points out the effect of the correction, namely, the
cancellation of the jumps between the micro-curves for energy,
and a shift of the curve for stress.
A decomposition of $\sigma_{11}$ into its different components is also
presented (bottom).
The open circles ($\circ$) represent the uncorrected values of isotropic
part of $\sigma_{11}$,
whereas the solid circles ($\bullet$) are the values corrected by the
{S.H.} technique.
The open diamonds ($\lozenge$) illustrate the values of the anisotropic
part of $\sigma_{11}$, that are not affected by the correction.}
\label{fig:sianiso}
\end{figure}

\begin{table}
\caption{List of the different sets of lattice constants
(expressed in atomic units) and the corresponding degree of the
polynomial used to fit the data.}
\label{tab:scales}
\begin{tabular}{cccccc}
&Scale &Points of the set &Number of points &Degree of the polynomial \\
\hline
&0.01 &10.00 to 10.50 &51 &3 &\\
&0.025 &10.00 to 10.50 &21 &3 &\\
&0.05 &9.90 to 10.60 &15 &3 &\\
&0.1 &9.50 to 11.00 &16 &3 &\\
&0.2 &9.00 to 11.60 &14 &4 &\\
&0.5 &7.00 to 13.00 &13 &8 &\\
\end{tabular}
\end{table}

\begin{table}
\caption{Equilibrium lattice constant calculated at different scales
($E_{cut}$=6 Ha with 2 special $k$-points in the IBZ)
from the energy curve before (Before) and after correction
with Francis-Payne technique (F-P)
or scaling hypothesis technique (S.H.), and
from the pressure curve before (Before) and after correction
with Froyen-Cohen technique (F-C)
or scaling hypothesis technique (S.H.).
The values are expressed in Bohr.
$\sigma$ is the standard deviation between the results obtained for the
different scales.}
\label{tab:ascales}
\begin{tabular}{ccc|cccc|cccc}
& & &\multicolumn{3}{c}{From energy} & &\multicolumn{3}{c}{From pressure}& \\
\hline
&Scale & &Before &F-P &S.H. & &Before &F-C &S.H. & \\
\hline
&0.01 & &10.1672 &10.2340 &10.2341 & &10.0301 &10.2234 &10.2245 & \\
&0.025 & &10.1678 &10.2349 &10.2351 & &10.0312 &10.2236 &10.2243 & \\
&0.05 & &10.1997 &10.2241 &10.2239 & &10.0308 &10.2241 &10.2245 & \\
&0.1 & &10.2136 &10.2266 &10.2262 & &10.0295 &10.2236 &10.2224 & \\
&0.2 & &10.2196 &10.2243 &10.2241 & &10.0272 &10.2241 &10.2249 & \\
&0.5 & &10.2287 &10.2188 &10.2188 & &10.0280 &10.2256 &10.2241 & \\
\hline
&$\mu $ & &10.1994 &10.2271 &10.2270
 & &10.0294 &10.2241 &10.2241 & \\
&$\sigma$ & &$2.6~10^{-2}$ &$6.2~10^{-3}$ &$6.3~10^{-3}$
 & &$1.6~10^{-3}$ &$8.1~10^{-4}$ &$8.8~10^{-4}$ & \\
\end{tabular}
\end{table}

\begin{table}
\caption{Equilibrium bulk modulus calculated at different scales
($E_{cut}$=6 Ha with 2 special $k$-points in the IBZ)
from the energy curve before (Before) and after correction
with Francis-Payne technique (F-P)
or scaling hypothesis technique (S.H.), and
from the pressure curve before (Before) and after correction
with Froyen-Cohen technique (F-C)
or scaling hypothesis technique (S.H.).
The values are expressed in Mbar.
$\sigma$ is the standard deviation between the results obtained for the
different scales.}
\label{tab:bscales}
\begin{tabular}{ccc|cccc|cccc}
& & &\multicolumn{3}{c}{From energy} & &\multicolumn{3}{c}{From pressure}& \\
\hline
&Scale & &Before &F-P &S.H. & &Before &F-C &S.H. & \\
\hline
&0.01 & &1.181 &0.8243 &0.8233 & &1.071 &0.9849 &0.9419 & \\
&0.025 & &1.079 &0.8390 &0.8368 & &1.088 &0.9823 &0.9417 & \\
&0.05 & &0.8326 &0.8860 &0.8859 & &1.123 &0.9560 &0.9426 & \\
&0.1 & &1.0030 &0.9709 &0.9710 & &1.127 &0.9482 &0.9412 & \\
&0.2 & &0.9546 &0.9630 &0.9601 & &1.115 &0.9333 &0.9285 & \\
&0.5 & &0.9539 &0.9485 &0.9481 & &1.108 &0.9321 &0.9411 & \\
\hline
&$\mu $ & &1.0007 &0.9053 &0.9042
 & &1.1053 &0.9561 &0.9395 & \\
&$\sigma$ & &$1.2~10^{-1}$ &$6.5~10^{-2}$ &$6.5~10^{-2}$
 & &$2.2~10^{-2}$ &$2.3~10^{-2}$ &$5.4~10^{-3}$ & \\
\end{tabular}
\end{table}

\begin{table}
\caption{Equilibrium lattice constant calculated for the different micro-curves
($E_{cut}$=6 Ha with 2 special $k$-points in the IBZ)
from the energy curve before (Before) and after correction
with Francis-Payne technique (F-P)
or scaling hypothesis technique (S.H.), and
from the pressure curve before (Before) and after correction
with Froyen-Cohen technique (F-C)
or scaling hypothesis technique (S.H.).
The values are expressed in Bohr.
$\sigma$ is the standard deviation between the results obtained for the
different scales.}
\label{tab:asegments}
\begin{tabular}{ccc|cccc|cccc}
& & &\multicolumn{3}{c}{From energy} & &\multicolumn{3}{c}{From pressure}& \\
\hline
&Micro-curve
 & &Before &F-P &S.H. & &Before &F-C &S.H. & \\
\hline
&1 & &10.0263 &10.2000 &10.2287 & &10.0263 &10.1996 &10.2279 & \\
&2 & &10.0380 &10.2138 &10.2262 & &10.0380 &10.2135 &10.2258 & \\
&3 & &10.0423 &10.2188 &10.2260 & &10.0425 &10.2188 &10.2260 & \\
&4 & &10.0537 &10.2247 &10.2241 & &10.0473 &10.2245 &10.2241 & \\
&5 & &10.0529 &10.2328 &10.2224 & &10.0544 &10.2328 &10.2226 & \\
&6 & &10.0577 &10.2400 &10.2216 & &10.0605 &10.2404 &10.2222 & \\
&7 & &10.0671 &10.2527 &10.2200 & &10.0711 &10.2534 &10.2213 & \\
\hline
&$\mu$ & &10.0482 &10.2261 &10.2241
 & &10.0485 &10.2261 &10.2241 &\\
&$\sigma$ & &$1.4~10^{-2}$ &$1.7~10^{-2}$ &$3.1~10^{-3}$
 & &$1.5~10^{-2}$ &$1.8~10^{-2}$ &$2.4~10^{-3}$ &\\
\end{tabular}
\end{table}

\begin{table}
\caption{Equilibrium bulk modulus calculated for the different micro-curves
($E_{cut}$=6 Ha with 2 special $k$-points in the IBZ)
from the energy curve before (Before) and after correction
with Francis-Payne technique (F-P)
or scaling hypothesis technique (S.H.), and
from the pressure curve before (Before) and after correction
with Froyen-Cohen technique (F-C)
or scaling hypothesis technique (S.H.).
The values are expressed in Mbar.
$\sigma$ is the standard deviation between the results obtained for the
differentscales.}
\label{tab:bsegments}
\begin{tabular}{ccc|cccc|cccc}
& & &\multicolumn{3}{c}{From energy} & &\multicolumn{3}{c}{From pressure}& \\
\hline
&Micro-curve
 & &Before &F-P &S.H. & &Before &F-C &S.H. & \\
\hline
&1 & &1.244 &1.080 &0.9178 & &1.244 &1.087 &0.9294 & \\
&2 & &1.226 &1.063 &0.9203 & &1.227 &1.069 &0.9278 & \\
&3 & &1.215 &1.062 &0.9253 & &1.220 &1.062 &0.9261 & \\
&4 & &1.278 &1.071 &0.9428 & &1.211 &1.056 &0.9281 & \\
&5 & &1.125 &1.044 &0.9260 & &1.201 &1.047 &0.9295 & \\
&6 & &1.165 &1.031 &0.9201 & &1.189 &1.037 &0.9285 & \\
&7 & &1.142 &1.012 &0.9143 & &1.171 &1.022 &0.9276 & \\
\hline
&$\mu$ & &1.1993 &1.0518 &0.9238
 & &1.2090 &1.0543 &0.9281 &\\
&$\sigma$ & &$5~10^{-2}$ &$2~10^{-2}$ &$1~10^{-2}$
 & &$2~10^{-2}$ &$2~10^{-2}$ &$1~10^{-3}$ &\\
\end{tabular}
\end{table}

\begin{table}
\caption{Equilibrium lattice constant calculated with different cut-off energy
at scale 0.1
(with 2 special $k$-points in the IBZ)
from the energy curve before (Before) and after correction
with Francis-Payne technique (F-P)
or scaling hypothesis technique (S.H.), and
from the pressure curve before (Before) and after correction
with Froyen-Cohen technique (F-C)
or scaling hypothesis technique (S.H.).
The values are expressed in Bohr.
The values between brackets indicate the relative error with respect to the
fully
converged value at 15 Ha, for the same technique.}
\label{tab:aconverge}
\begin{tabular}{ccc|cccc|cccc}
& & &\multicolumn{3}{c}{From energy} & &\multicolumn{3}{c}{From pressure} & \\
\hline
&$E_{cut}$
 & &Before &F-P &S.H. & &Before &F-C &S.H. & \\
\hline
&3 & &10.1121 &10.0935 &10.0935 & &9.6899 &10.0856 &10.0854 & \\
& & &(0.8\%) &(0.9\%) &(0.9\%) & &(5\%) &(1\%) &(1\%) & \\
&6 & &10.2135 &10.2266 &10.2262 & &10.0295 &10.2236 &10.224 & \\
& & &(0.2\%) &(0.4\%) &(0.4\%) & &(2\%) &(0.4\%) &(0.4\%) & \\
&10 & &10.1897 &10.1894 &10.1894 & &10.1771 &10.1867 &10.1869 & \\
& & &($\ll$0.1\%) &($\ll$0.1\%) &($\ll$0.1\%)
 & &($\ll$0.1\%) &($\ll$0.1\%) &($\ll$0.1\%)& \\
&15 & &10.1886 &10.1888 &10.1888 & &10.1716 &10.1856 &10.1856 & \\
\end{tabular}
\end{table}

\begin{table}
\caption{Equilibrium bulk modulus calculated with different cut-off energy at
scale 0.1
(with 2 special $k$-points in the IBZ)
from the energy curve before (Before) and after correction
with Francis-Payne technique (F-P)
or scaling hypothesis technique (S.H.), and
from the pressure curve before (Before) and after correction
with Froyen-Cohen technique (F-C)
or scaling hypothesis technique (S.H.).
The values are expressed in Mbar.
The values between brackets indicate the relative error with respect to the
fully
converged value at 15 Ha, for the same technique.}
\label{tab:bconverge}
\begin{tabular}{ccc|cccc|cccc}
& & &\multicolumn{3}{c}{From energy} & &\multicolumn{3}{c}{From pressure}& \\
\hline
&$E_{cut}$
 & &Before &F-P &S.H. & &Before &F-C &S.H. & \\
\hline
&3 & &1.126 &1.169 &1.169 & &1.620 &1.154 &1.154 & \\
& & &(14\%) &(18\%) &(18\%) & &(64\%) &(19\%) &(19\%) & \\
&6 & &1.003 &0.9709 &0.9710 & &1.1270 &0.9482 &0.9412 & \\
& & &(2\%) &(2\%) &(2\%) & &(15\%) &(2\%) &(3\%) & \\
&10 & &0.9889 &0.9902 &0.9902 & &0.9805 &0.9712 &0.9714 & \\
& & &(0.2\%) &(0.3\%) &(0.3\%) & &(0.2\%) &(0.2\%) &(0.2\%)& \\
&15 & &0.9875 &0.9872 &0.9872 & &0.9826 &0.9692 &0.9692 & \\
\end{tabular}
\end{table}


\begin{thebibliography}{10}
\bibitem{Note_kpoints}
Since the number of plane waves varies with the electronic wavevector $k$, when
  we note $\overline{N}_{PW}$ we refer to its average on the special $k$-point
  set. See Appendix \ref{app:meanonk} for more information.
\bibitem{Note_npw}
For sake of brevity, we will not recall in what follows that
  $\overline{N}_{PW}$ is a mean number, it will only be suggested by the bar.
\bibitem{Note_hypecut}
This expression had already been suggested by Francis and Payne
  (ref.\cite{Francis-Payne90} see bottom of page 4400).
\bibitem{Note_noise}
Strictly speaking, the jumps between the micro-curves cannot be called
  ``noise'', since their origin is deterministic, and not statistical. But the
  analysis that follows uses the tools of statistics, since in any realistic
  application the ragged curve (shown in Fig.~\ref{fig:segments}) will be
  sampled nearly at random by a few points (as in Sec.~\ref{sec:applic}), and
  the jumps will have an effect similiar to a noise affecting a smooth curve.
\end{thebibliography}
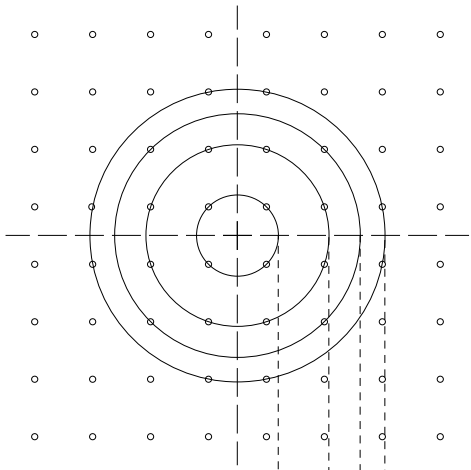
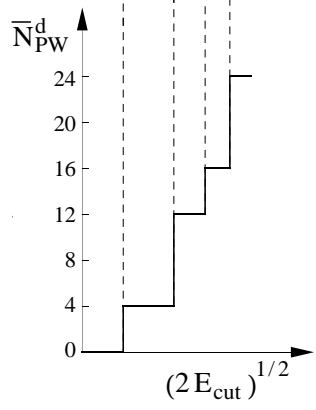
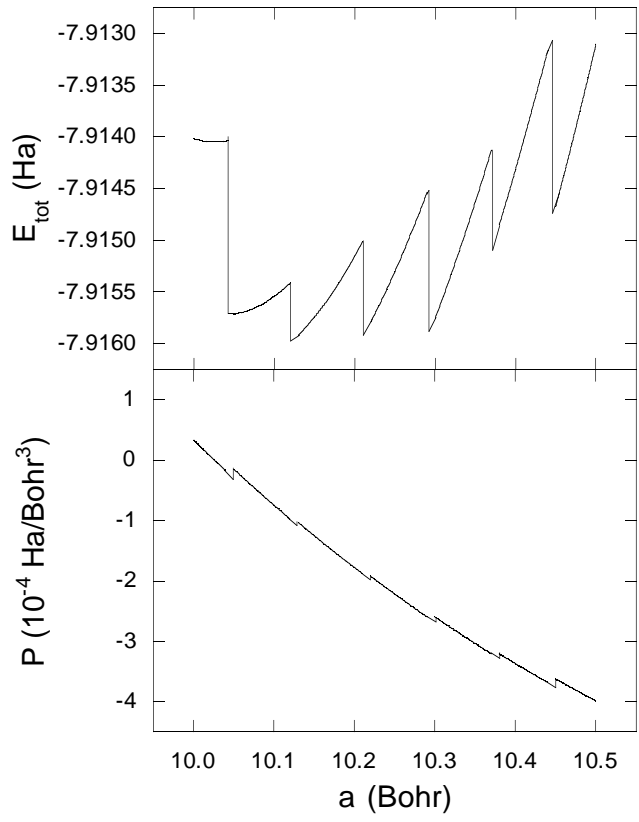
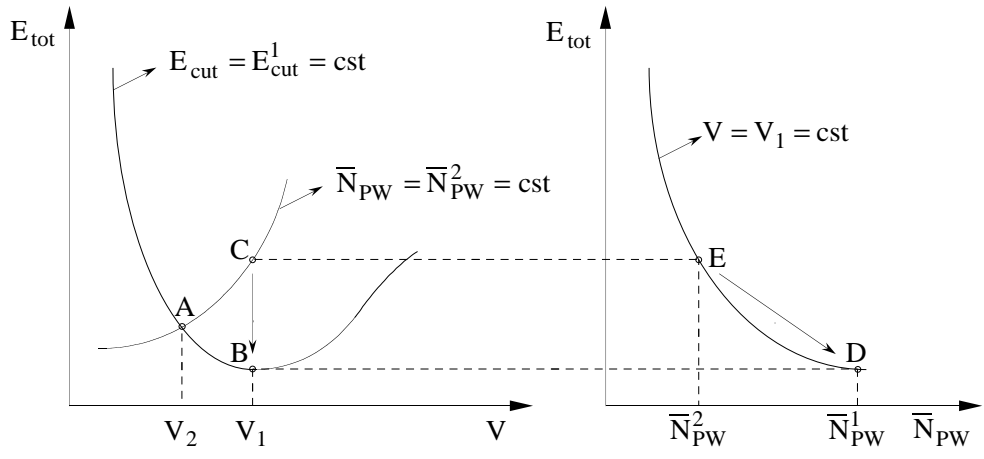
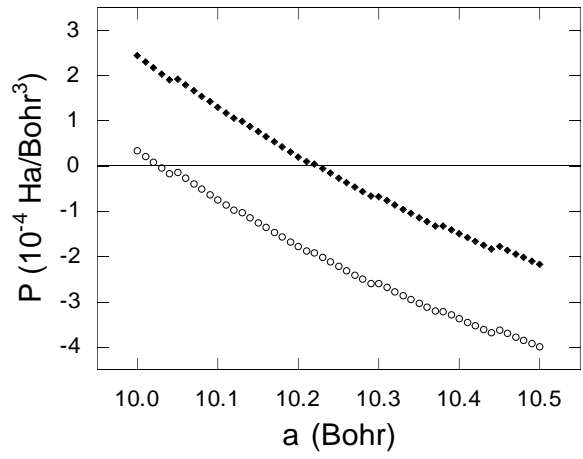
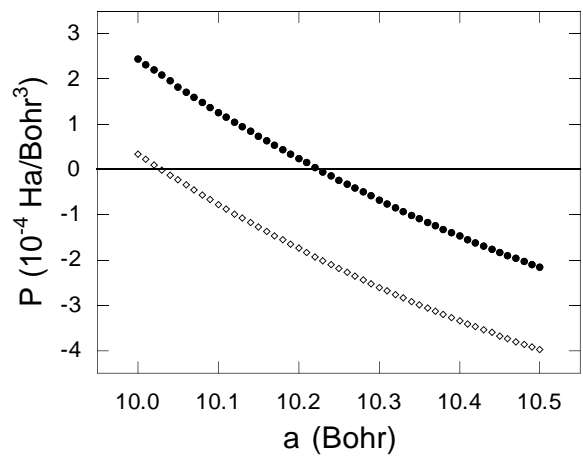
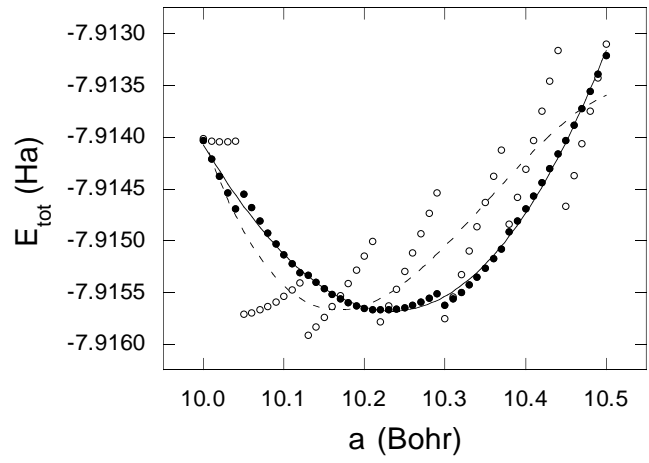
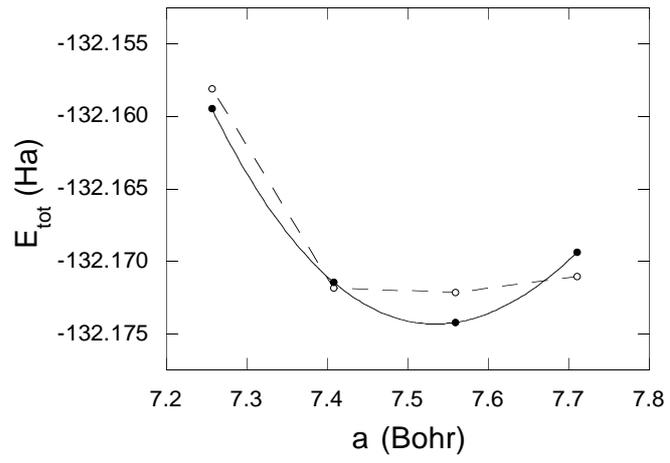
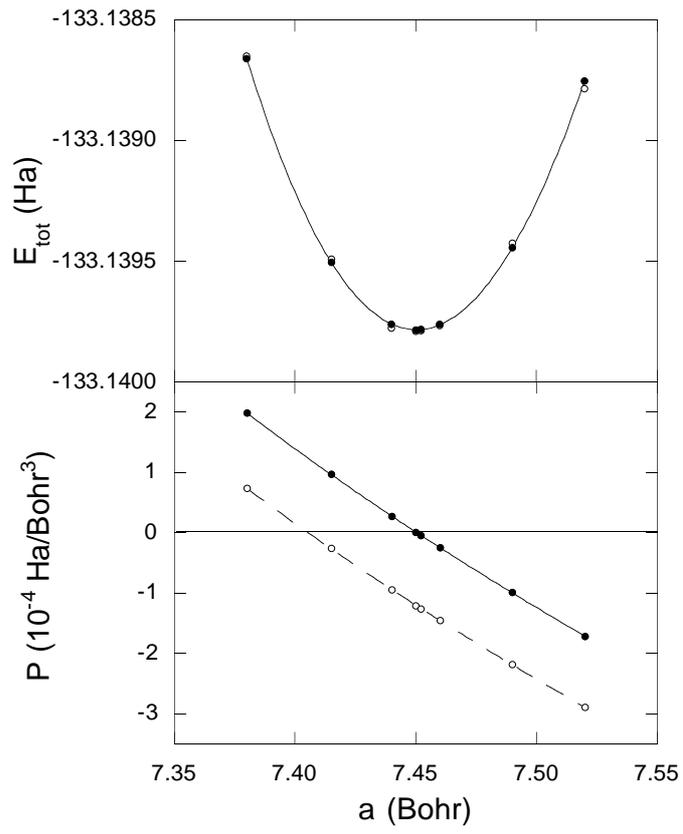
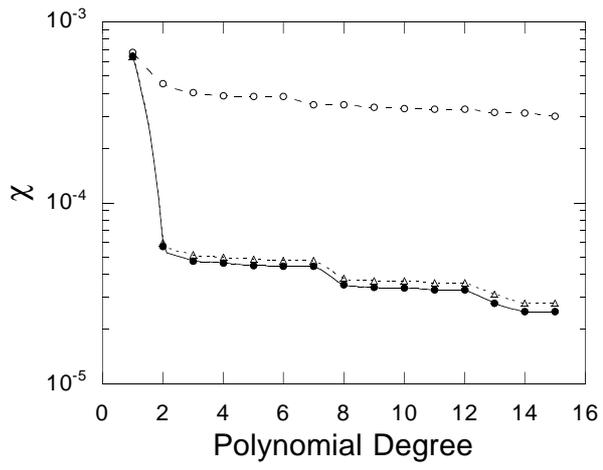
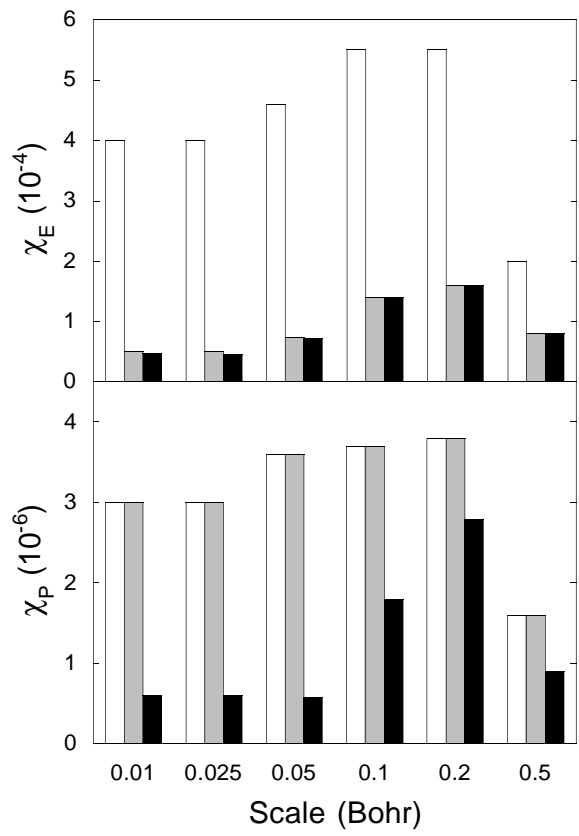
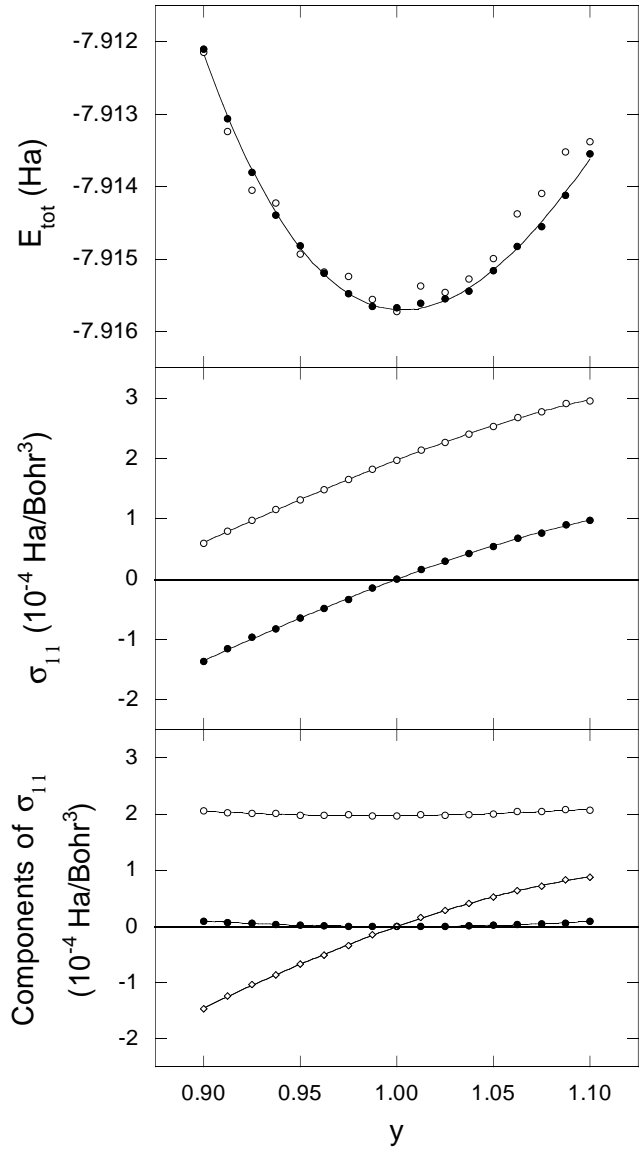
\end{document}